\documentclass{aa}  
\usepackage{subfigure}
\usepackage{graphicx}
\usepackage{natbib}
\bibpunct{(}{)}{;}{a}{}{,}
\usepackage{txfonts}
\begin{document}
   \title{H$_2$CO and CH$_3$OH maps of the Orion Bar photodissociation region\thanks{Based on observations carried out with the IRAM-30m telescope and the Plateau de Bure interferometer. IRAM is supported by INSU/CNRS (France), MPG (Germany) and IGN (Spain).}}

   \subtitle{}

   \author{S. Leurini\inst{1,2}, B. Parise\inst{2}, P. Schilke\inst{2,3}, J. Pety\inst{4}, R. Rolffs\inst{2}} 

   \offprints{S. Leurini}

   \institute{ESO, Karl-Scharzschild-Strasse 2, D-85748, Garching-bei-M\"unchen, Germany\\\email{sleurini@mpifr.de}
\and
Max-Planck-Institut f\"ur Radioastronomie,
              Auf dem H\"ugel 69, 53121 Bonn, Germany
\and
Physikalisches Institut, Universit\"at zu K\"oln, Z\"ulpicher Str. 77, 50937 K\"oln, Germany
\and 
Institut de Radioastronomie Millim\'etrique, 300 Rue de la Piscine, 38406 Saint Martin d'H\`eres, France}
   \date{\today}

 \abstract{A previous analysis of methanol and formaldehyde towards
the Orion Bar concluded that the two molecular species may trace
different physical components, methanol the clumpy material,
 and formaldehyde the interclump medium.} 
{To verify this
hypothesis, we performed multi-line mapping observations of the two
molecules to study their spatial distributions.}  
{The observations
were performed with the IRAM-30m telescope at 218 and 241~GHz, with an
angular resolution of $\sim 11''$.  Additional data for H$_2$CO from
the Plateau de Bure array are also discussed. The data were analysed
using an LVG approach.}  
{Both molecules are detected in our single-dish data. Our data show that CH$_3$OH peaks towards the clumps of the
Bar, but its intensity decreases below the detection threshold in the
interclump material. When averaging over a large region of the interclump medium, the strongest CH$_3$OH line is detected with a peak intensity of 
$\sim 0.06$~K. 
Formaldehyde also peaks on the clumps, but it is
also detected  in the interclump gas.}  
{We verified that the weak intensity of CH$_3$OH in the interclump medium is not caused by the
different excitation conditions of the interclump material, but
reflects a decrease in the column density of methanol.
 The abundance of CH$_3$OH relative to H$_2$CO decreases by at least
one order of magnitude from the dense clumps to the interclump
medium.}  
\authorrunning{S. Leurini et al.}
\titlerunning{H$_2$CO and CH$_3$OH maps of the Orion Bar}
   \maketitle
%

\section{Introduction}\label{intro}
Given its proximity and nearly edge-on orientation, the so-called
Orion Bar is one of the clearest examples of a photon-dominated region (PDR).
For this reason, the Orion Bar has been extensively observed in  past decades
to test theoretical models of PDR structure, chemistry, and
energetics. Its nearly edge-on orientation make direct observations
of the gas stratification possible, from ionised gas to neutral atomic gas to
molecular gas as a function of the increasing distance from the
ionisation source \citep[e.g.,][]{1993Sci...262...86T,1996A&A...313..633V,2009arXiv0902.1433V}. These studies show that the molecular
gas  consists of clumpy molecular cores
embedded in an interclump gas. While the clumps have densities of
several 10$^6$~cm$^{-3}$ \citep{2000ApJ...540..886Y,2003ApJ...597L.145L}, the interclump
material has lower densities
\citep[10$^4$--$10^5$~cm$^{-3}$,][]{1995A&A...294..792H,2000ApJ...540..886Y}. Such observational results are successfully 
reproduced by theoretical works  \citep[e.g.,][]{2002ApJ...573..215G}.

Observations of  methanol (CH$_3$OH) and formaldehyde (H$_2$CO) in the
Orion Bar have been reported in the past
\citep{1995A&A...294..792H,1995A&A...303..541J,2006A&A...454L..47L}. \citet{2006A&A...454L..47L}
studied the excitation of both molecular species in the line of sight
of one molecular clump through a multi-line analysis at 290~GHz. Their
findings (high density and relatively low temperature for CH$_3$OH, high temperature, 
and relatively low density for H$_2$CO) suggest that
methanol and formaldehyde do not trace the same material, but the first is
found in the dense gas associated with the clumps, the
second in the warmer and less dense gas of the interclump material.
Since both molecular species can efficiently form on grain surfaces
\citep{2004ApJ...614.1124H}, the authors suggested that
photodissociation of methanol to form formaldehyde
\citep{2000A&AS..146..157L} takes place in the interclump medium,
while the high density shields CH$_3$OH in the clumps and prevents its
photodissociation. Whereas grain surface reactions are the only viable route to methanol, 
formaldehyde can also form in the gas phase via the reaction CH$_3$+O \citep[e.g., ][]{2000A&AS..146..157L}, 
which could contribute to the interclump abundance of H$_2$CO.

\begin{figure*}
\centering
\subfigure[]{\includegraphics[width=16cm]{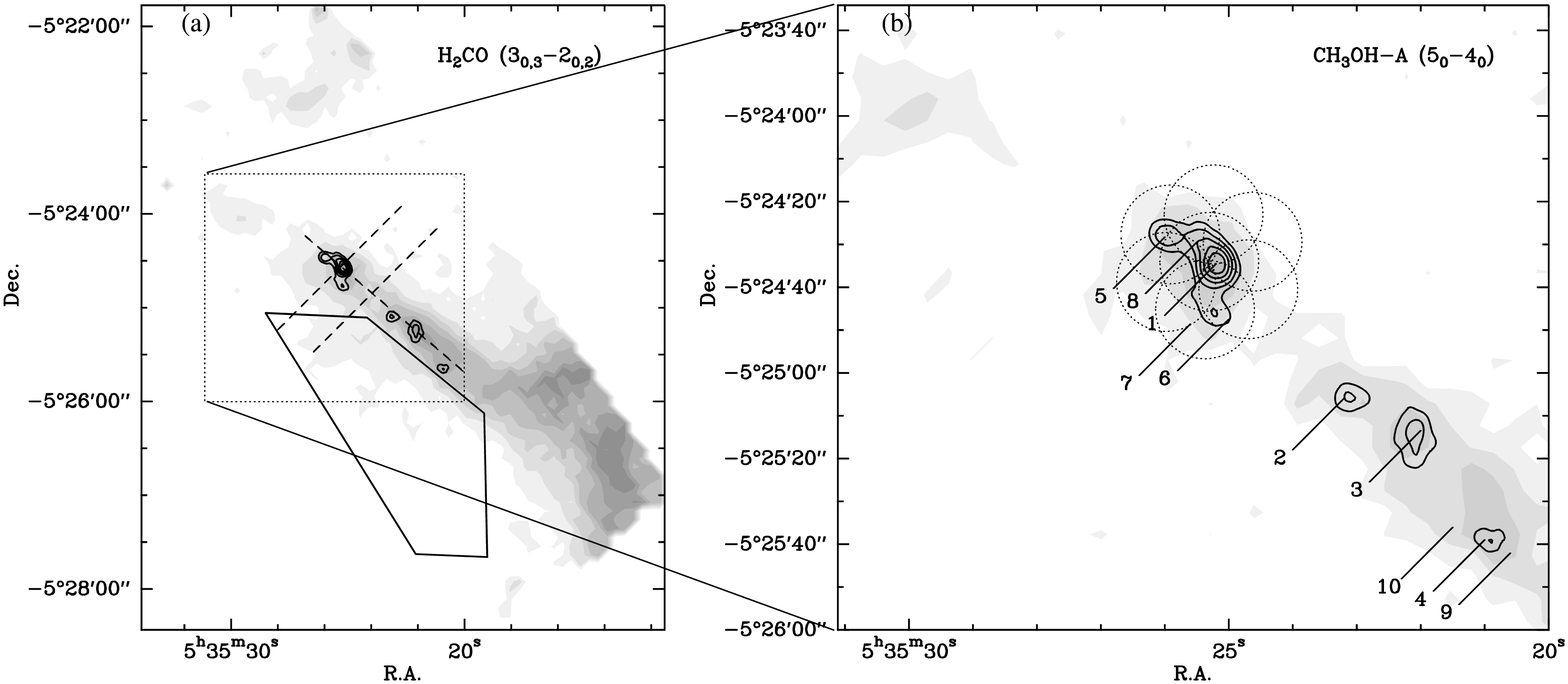}}
\caption{In grey scale, the integrated intensity of the H$_2$CO
($3_{0,3}-2_{0,2})$ line (left) and of the CH$_3$OH-$A$ $(5_0-4_0)$
transition (right)  observed with the IRAM-30m telescope. Black 
contours (4, 8, and 12 times 0.032
Jy~beam$^{-1}$~km~s$^{-1}$) show the distribution of the H$^{13}$CN
(1-0) emission observed with the Plateau de Bure interferometer
\citep{2003ApJ...597L.145L}.  For H$_2$CO the contours start from
1.8~K~km~s$^{-1}$ in steps of 1.8 (equal to 3$\sigma$), for CH$_3$OH
from 0.3~K~km~s$^{-1}$ in steps of 0.3 (equal to 3$\sigma$).  In the left panel, 
the
dashed lines indicate the strips used for Fig.~\ref{strip-ra}; the
solid lines outline the regions used to derive the average H$_2$CO and
CH$_3$OH emission in the interclump material (Figs.~\ref{h2co-inter}
and ~\ref{ch3oh-inter}); the dotted lines mark the region shown in the
right panel. In the right panel, the
numbers label the positions of the clumps detected in H$^{13}$CN; the dotted circles outline the mosaic of seven fields observed with the PdBI 
(see \S~\ref{pdbi}).}\label{integrated}
\end{figure*}

However, the study of \citet{2006A&A...454L..47L} on CH$_3$OH and
H$_2$CO is based on low angular resolution ($\sim 20''$), single-pointing data. 
To confirm that the two species indeed trace different
media, we mapped a large area of the Orion Bar in formaldehyde and
methanol at 218 and 241~GHz, respectively, with an angular resolution
of $\sim 11''$. We also analysed existing interferometric data  for the
218~GHz formaldehyde lines, to compare the observed fluxes from single-dish and interferometer and derive constraints on the size of the
emitting region.  This work represents a first observational
 attempt to study the
distribution of CH$_3$OH and H$_2$CO in the Orion Bar
and to verify their different origin.

\section{Observations}\label{par-obs}
\subsection{IRAM-30m telescope}
The observations were performed in February 
and March
 2007 using the HERA multi-beam receiver \citep{2004A&A...423.1171S}
at the IRAM-30m telescope.  The HERA1 and HERA2 pixels were tuned to
241.850~GHz in lower side band (LSB), to detect the CH$_3$OH ($5_k-4_k$) band. On February 23 and 24, the HERA2 receivers were
tuned to 218.349~GHz (LSB) to detect the H$_2$CO ($3_{K_a,K_c}-2_{K_a,K_c-1}$) band. The backend used was VESPA with a bandwidth of
160~MHz and a resolution of 0.3125~MHz for the CH$_3$OH setup
(corresponding to a velocity resolution of $\sim 0.4$~km~s$^{-1}$),
160~MHz and 1.25~MHz for the H$_2$CO setup (corresponding to $\sim
1.7$~km~s$^{-1}$). In this way, the CH$_3$OH ($5_k-4_k$) band is
fully covered with the only exception being the $k=-1,0$-$E$ lines. The
H$_2$CO setup covers the $3_{0,3}-2_{0,2}$ and $3_{2,2}-2_{2,1}$ lines, while the $3_{2,1}-2_{2,0}$ transition lies outside the observed frequency range.

We used the on-the-fly mode with a rotation of the multi-beam system
of 9.7$^\circ$, to ensure a Nyquist sampling between the rows.  The
reference position used as the centre of the map was 
$\alpha_{2000}=05^h35^m25^s.3, \delta_{2000}=-05^\circ24'34''.0$,
corresponding to the ``Orion Bar (HCN)'' position of
\citet{2001A&A...372..291S}, the most massive clump seen in H$^{13}$CN
\citep{2003ApJ...597L.145L}, as well as the target of the spectral
survey of \citet{2006A&A...454L..47L}. The OFF position was
chosen to be $(500'', 0'')$ from the centre of the map. 

The pointing was checked on several continuum sources
and found to be accurate within $2''$.
The observations were done
under varying weather conditions, and the T$_{\rm sys}$ span the range
540--1200~K for HERA1 and 700--1600~K for HERA2. The half-power beam width of the IRAM-30m telescope at 218~GHz is $\sim 11''$ 
and  $\sim10''$ at 242~GHz. We 
used  a beam efficiency  of 0.55 at 218~GHz, and of 0.50 at 242~GHz to convert antenna
temperatures T$^\star_{\rm a}$ into main-beam temperatures T$_{\rm mb}$\footnote{http://www.iram.es/IRAMES/telescope.html}.

The region covered by our observations is shown in Fig.~\ref{integrated}.

\subsection{Plateau de Bure interferometer}\label{pdbi}

\begin{figure}
    \centering{}
     \includegraphics[height=0.7\hsize,angle=270]{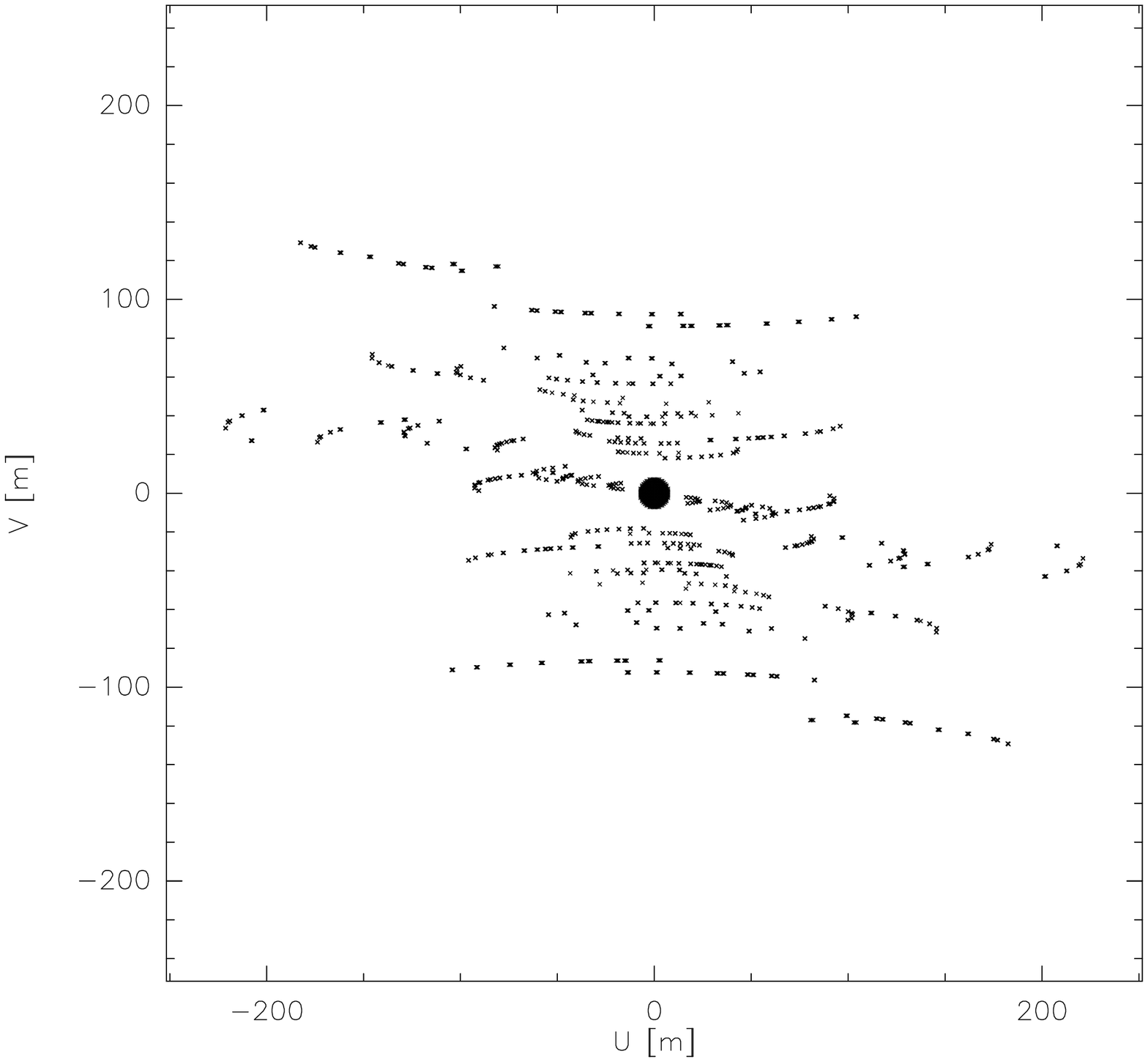}
     \includegraphics[height=0.72\hsize,angle=270]{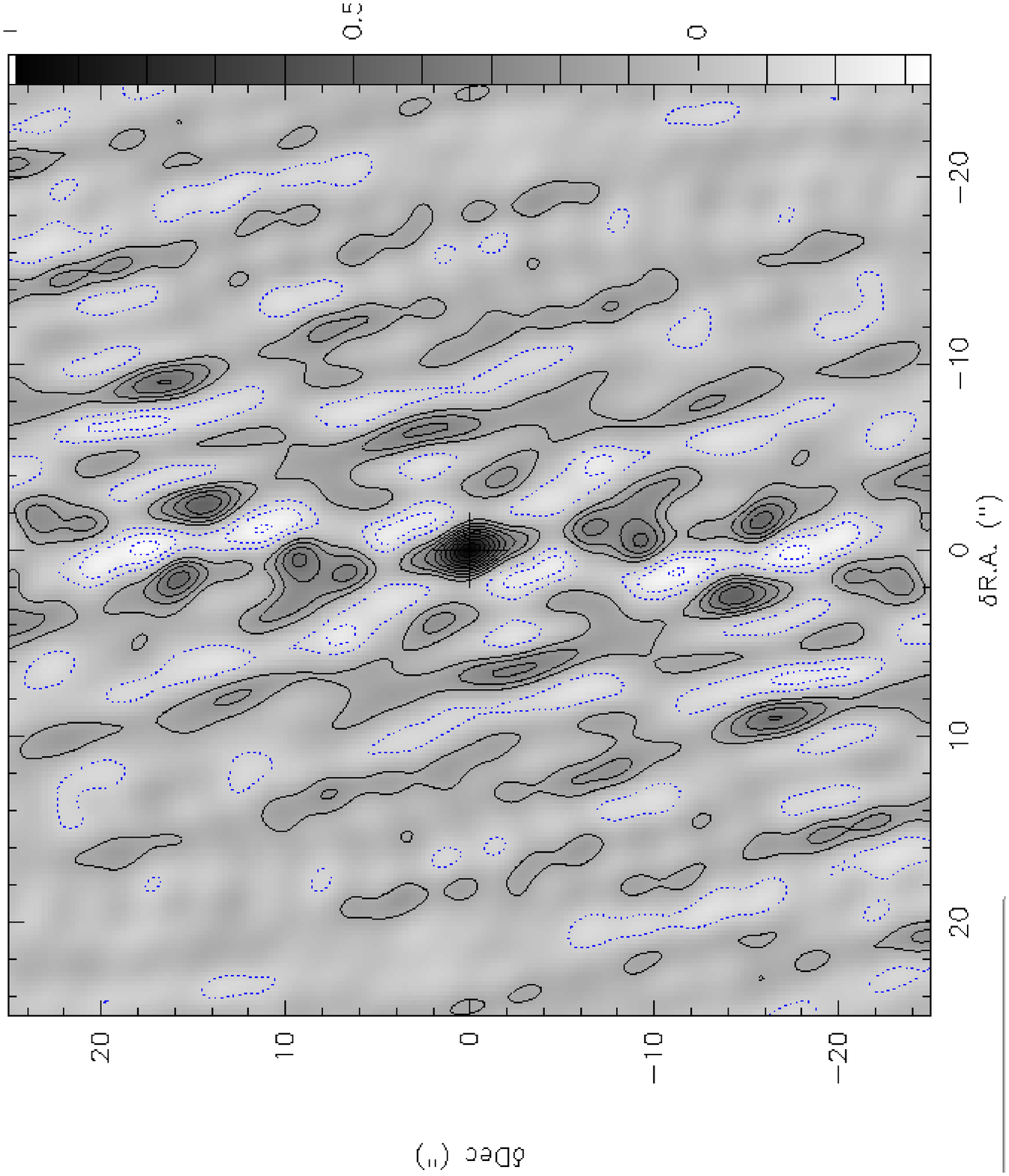}
    \caption{\emph{Top:} $uv$ coverage for one field of the hybrid
      30m+PdBI mosaic. \emph{Bottom:} Associated dirty beam. Please note
      that the dirty beam has many secondary side-lobes as high as 40\% of
      the main lobe.}\label{fig_beam}
\end{figure}

 We observed the H$_2$CO $  3_{0,3}-2_{0,2}$ line at 218.2~GHz with the Plateau de Bure
interferometer in March, April, and December 2004. We observed a 7-field
mosaic centred on $\alpha_{2000} = 05^h35^m25.280^s$, $\delta_{2000} =
-05^\circ24'39''.300$. The field positions followed a compact hexagonal
pattern to ensure Nyquist sampling in all directions. The imaged
field-of-view is almost a circle with a radius of $22.5''$. On March
28 and 29, the six antennas were used in the 6 Cp configuration with 3 and 1.5~mm precipitable water vapour, 
which translated into  system
temperatures of 450~K and 250 K. On April 22, the six antennas were used in
the more compact 6Dp configuration with 6 to 10~mm PWV,
leading to a system temperature of about 1000 K. On December 12, the six
antennas were used in the 6Cp configuration, with 3.5~mm PWV,
leading to a system temperature of about 700 K. Taking the
time for calibration and data filtering into account, this translates into an 
\emph{on--source} integration time of useful data of 5.3~hours for a full
6-antenna array. The typical 1 mm resolution is $2.4''$. We used the 30m
data described above to produce the missing short spacings.

The data processing was done with the GILDAS\footnote{See
  \texttt{http://www.iram.fr/IRAMFR/GILDAS} for more information about the
  GILDAS softwares.} software suite~\citep{pety05}. Standard calibration
methods implemented in the GILDAS/CLIC program were applied using close
bright quasars (0528$+$134 and 0607$-$157) as phase and amplitude
calibrators. The absolute flux was calibrated using simultaneous
measurements of the PdBI primary flux calibrator MWC349. $uv$ tables were
then produced at a relatively coarse velocity resolution of 1.7 km/s.

 All other processing was performed with the GILDAS/MAPPING software.  The
single-dish map from the IRAM-30m was used to create the short-spacing
pseudo-visibilities unsampled by the Plateau de Bure
interferometer~\citep{rodriguez08}.  These were then merged with the
interferometric observations. Each mosaic field was imaged independently, 
and a dirty mosaic was built through a linear combination of these dirty
images~\citep{gueth95}. The dirty mosaic was then deconvolved using the
standard H\"ogbom CLEAN algorithm. While all these procedures are standard
and usually easy to use in the GILDAS/MAPPING software, we had to take
special precautions with this data set because the short usable integration
time at PdBI implied 1) a low signal-to-noise ratio of the interferometric
data and 2) dirty beams for each field with large secondary side-lobes (see
Fig.~\ref{fig_beam}). As the 30m data show that the extended signal fills most
of the field-of-view observed with the interferometer, we subtracted  
the spectrum averaged over the observed field-of-view from the 30m data before any
processing and we added this averaged spectrum again after the deconvolution
of the 30m+PdBI hybrid data set to recover the correct flux scale. This
simplifies  the deconvolution considerably by the usual CLEAN algorithms
(\emph{i.e.}\ much fewer CLEAN components are needed) because it avoids the
deconvolution of extended uniform intensities. Second, we tapered the
visibility weights with an axis-symmetrical Gaussian of 80m {\it FWHM} to get a
more symmetrical beam (going from $4.1''\times1.2''$ for natural weighting
to $3.3''\times 1.8''$ after tapering) and to improve the signal-to-noise
ratio for the extended structures.  The final
noise rms measured at the centred of the mosaic is about 0.3~K in channels
of 1.7 km/s width. This leads to a maximum signal-to-noise ratio of 18 for
the hybrid data cube. However, the produced data cube has its brightness
dynamical range limited by the poor dirty beam. The cleaned data cube was finally scaled
from Jy/beam to T$_{\mbox{mb}}$ temperature scale using the synthesised beam
size.

\section{Observational results}

\begin{figure*}
\centering
\subfigure[]{\label{h2co}       \includegraphics[bb= 189 16 552 682, angle=-90, width=12cm,clip]{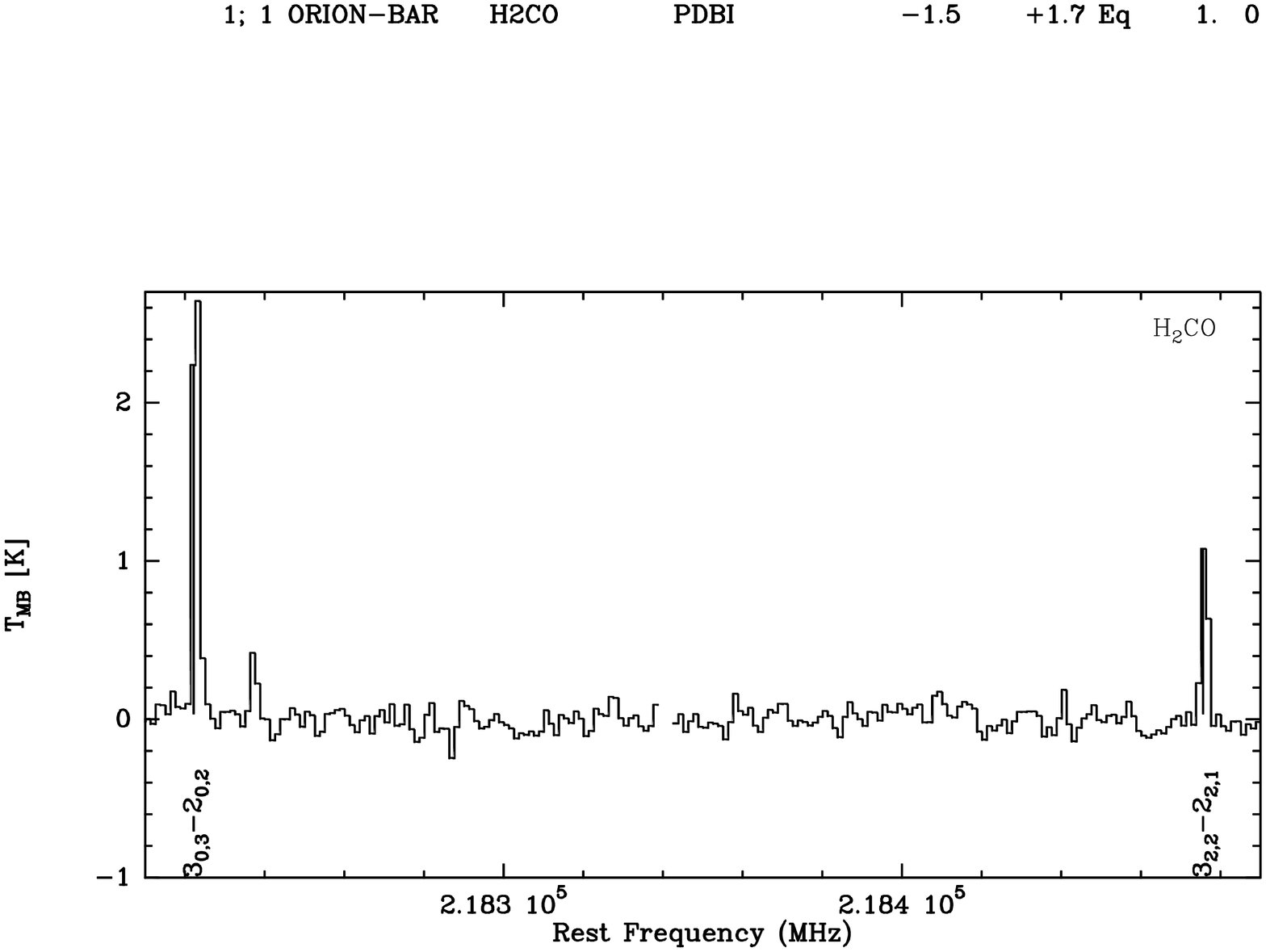}}
\subfigure[]{\label{h2co-inter}\includegraphics[bb= 189 16 552 682, angle=-90, width=12cm,clip]{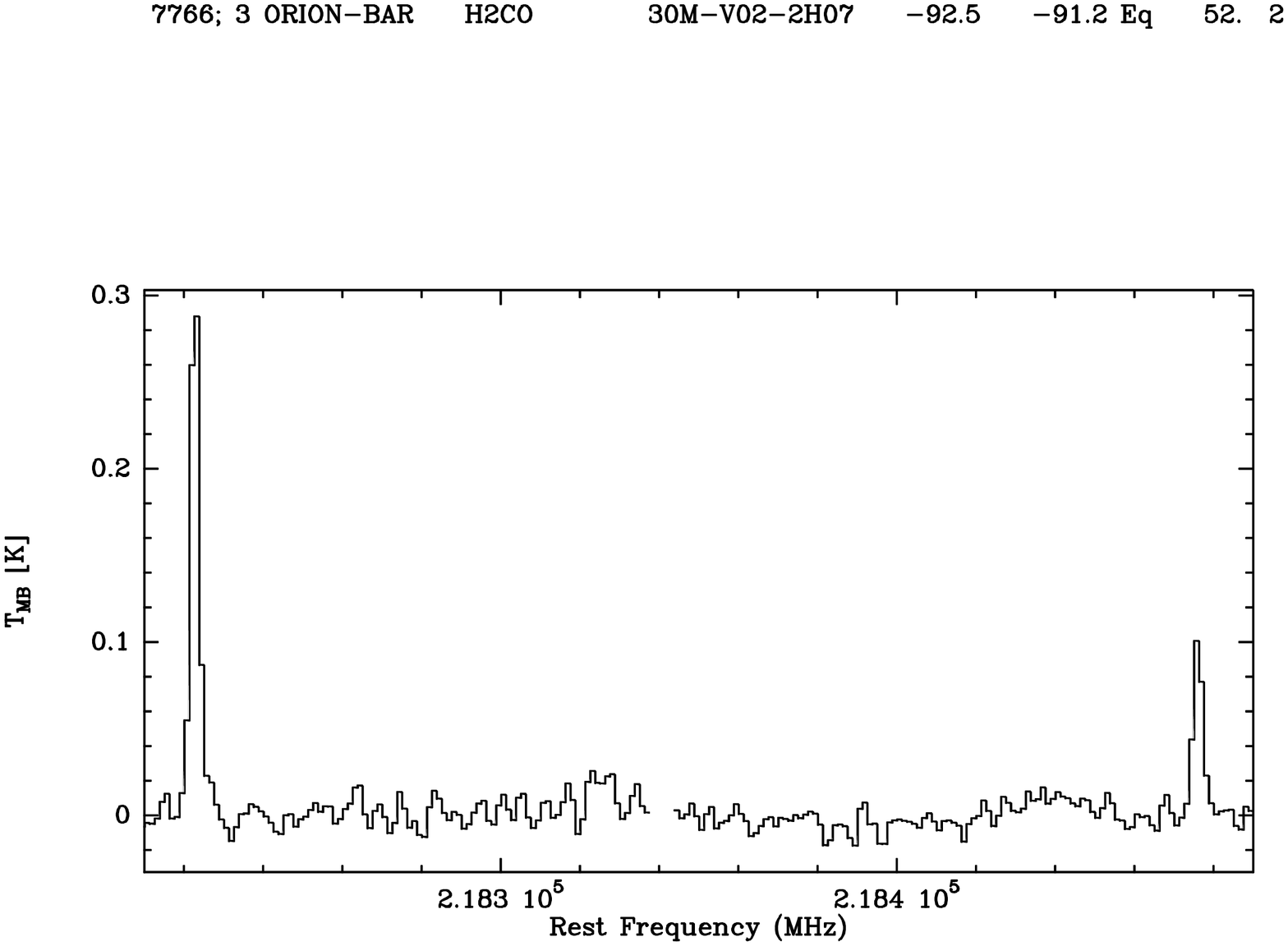}}
\caption{Spectrum of the H$_2$CO ($3_{K_a,K_c}-2_{K_a,K_c-1}$) band towards clump 1 (top) and averaged over the 
interclump medium (bottom) from the IRAM-30m telescope.}\label{spectra-h2co}
\end{figure*}
\begin{figure*}
\centering
\subfigure[]{\label{ch3oh}      \includegraphics[bb= 189 16 552 682, angle=-90, width=11.5cm,clip]{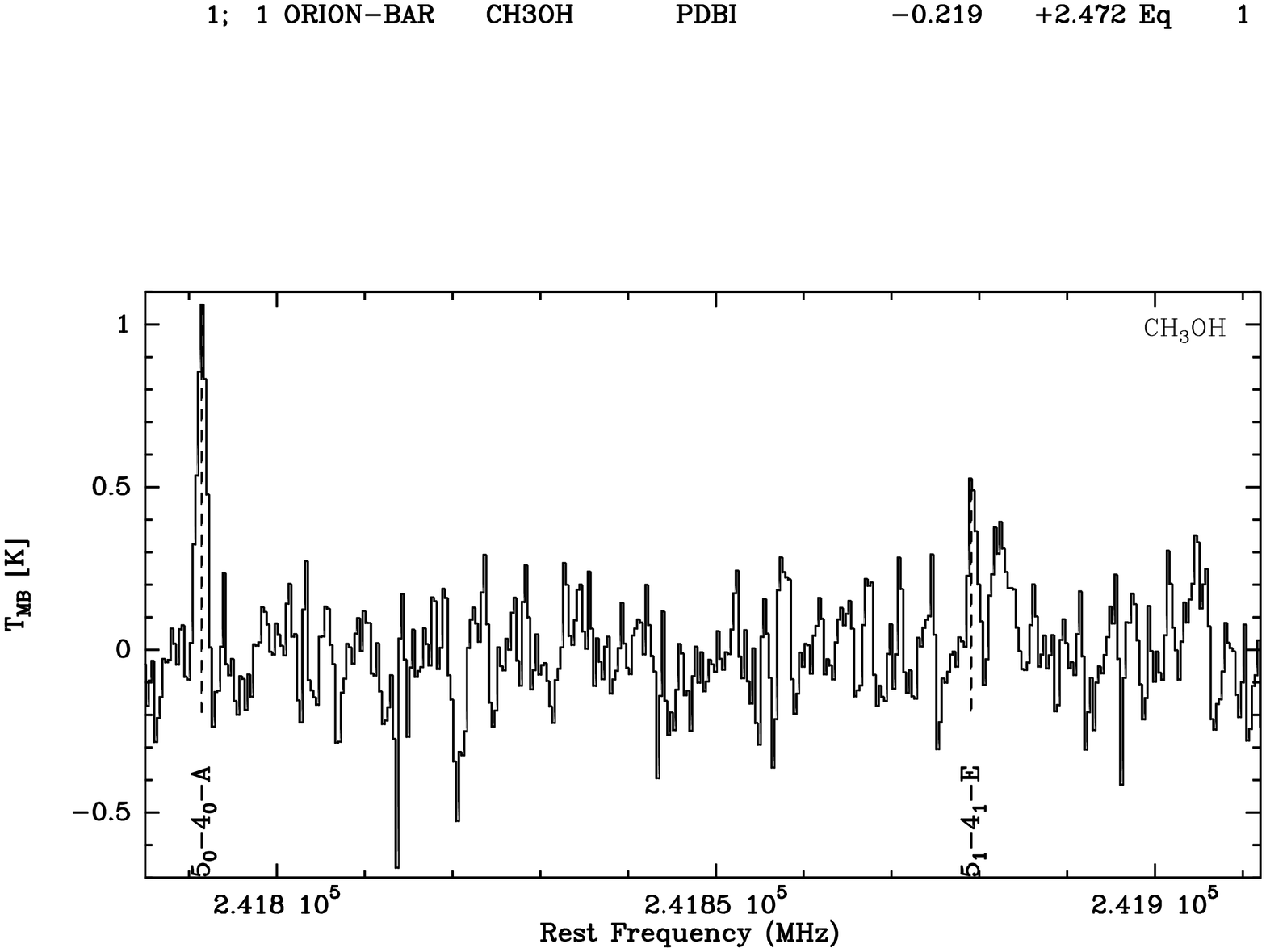}  }
\subfigure[]{\label{ch3oh-smooth}\includegraphics[bb= 185 15 551 682, clip,angle=-90, width=11.5cm]{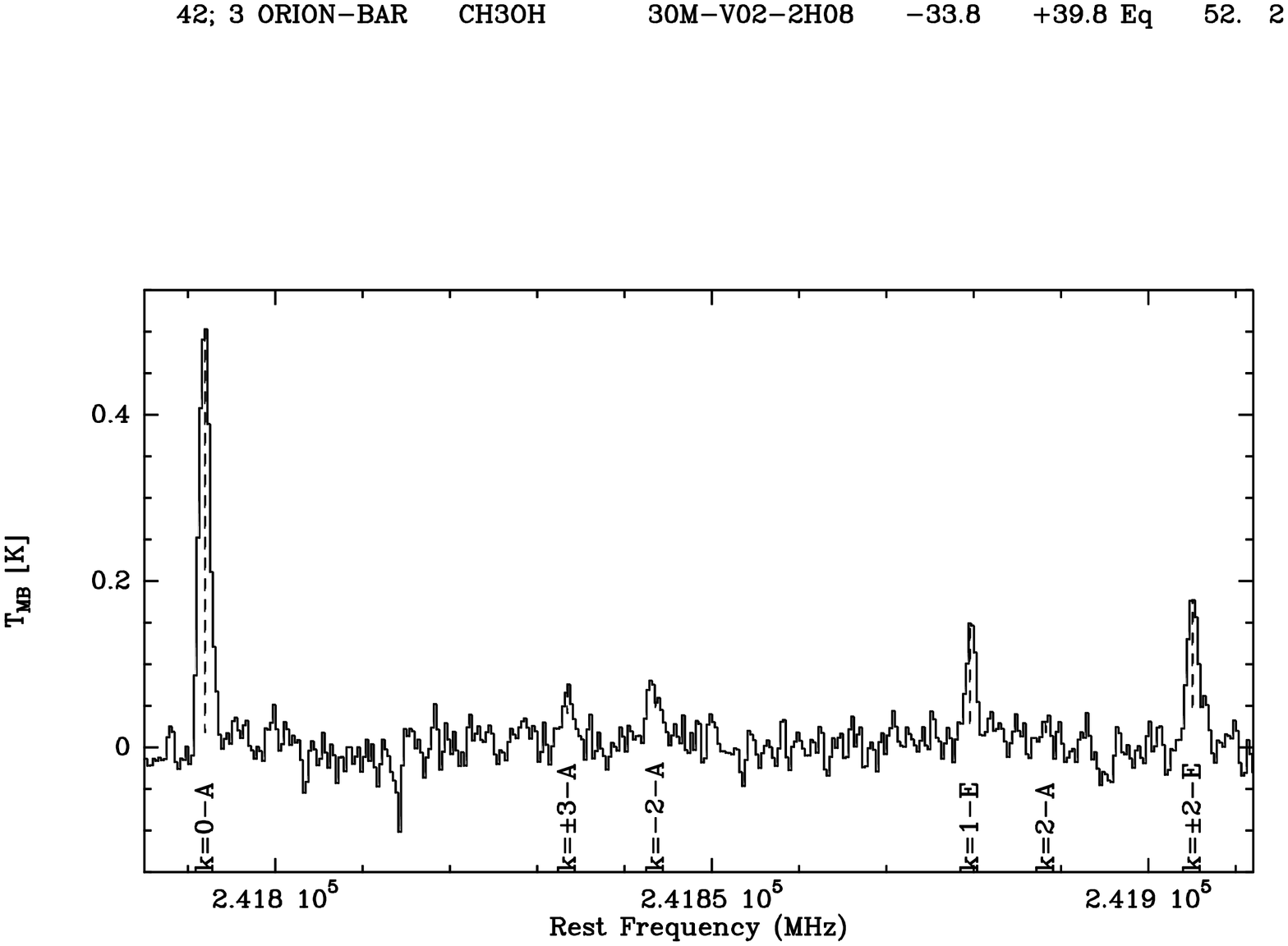}}
\subfigure[]{\label{ch3oh-inter}\includegraphics[bb= 189 16 552 682, angle=-90, width=11.5cm,clip]{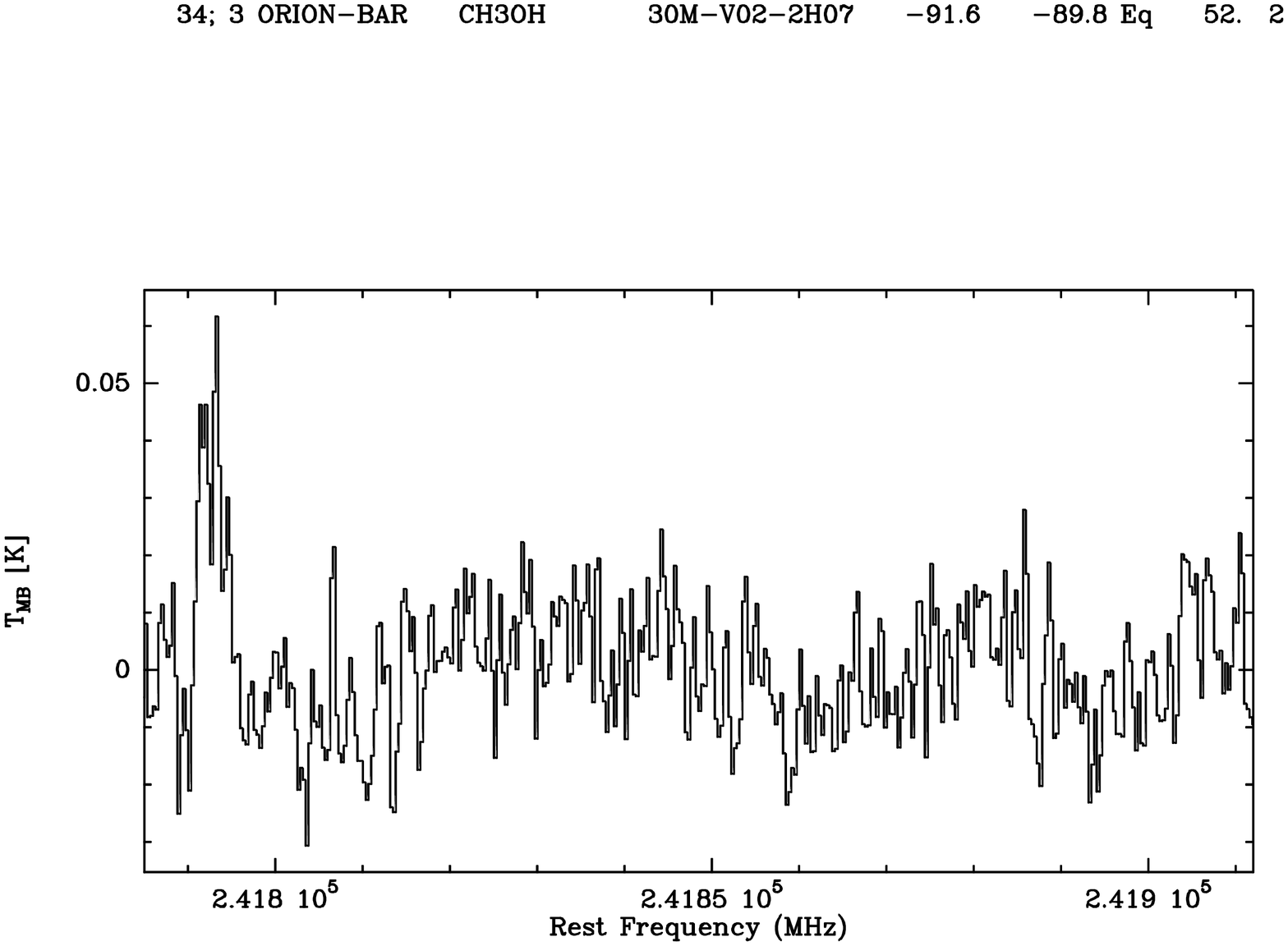}}
\caption{Spectrum of the CH$_3$OH ($5_k-4_k$) band towards clump 1  at a resolution of $\sim 10~\arcsec$ (top), smoothed to a resolution of $\sim 20~\arcsec$ (middle), and  averaged over the interclump medium (bottom). Data are from the IRAM-30m 
telescope.}\label{spectra-ch3oh}
\end{figure*}

\subsection{single-dish data}\label{30m}

Figures~\ref{h2co} and \ref{ch3oh} show the spectra  at 218.4 and 241.8~GHz 
towards the central position of the map.  Both formaldehyde transitions are detected. 
 Only the $5_0-4_0$-$A$ and $5_1-4_1$-$E$ methanol transitions are detected at the original resolution
of  $\sim 10\arcsec$. However, when increasing the signal-to-noise ratio by 
smoothing to a lower resolution of $\sim 20\arcsec$, CH$_3$OH transitions 
 with upper level energies lower than 85~K are also detected (Fig.~\ref{ch3oh-smooth}).  
The detected lines of both molecular species are reported in Table~\ref{lineid}. 

\begin{table}
\caption{Detected methanol and formaldehyde transitions in the IRAM-30m telescope data}\label{lineid}
\begin{tabular}{lcrcc}
\hline\hline
\multicolumn{1}{c}{line}&\multicolumn{1}{c}{$\nu$} &\multicolumn{1}{c}{E$_{up}$}&\multicolumn{1}{c}{clump 1}&\multicolumn{1}{c}{interclump}\\
&\multicolumn{1}{c}{[MHz]}&\multicolumn{1}{c}{[K]}&\\

 H$_2$CO        $  3_{0,3}   -      2_{0,2}      $    &        218222 &  21.0     & Y$^a$ & Y\\
 H$_2$CO        $  3_{2,2}   -      2_{1,1}      $    &        218475 &  68.0     & Y$^a$ & Y\\
 CH$_3$OH-$A$   $  5_0       -      4_0          $    &        241791 &  34.8     & Y$^a$ & Y\\
 CH$_3$OH-$A$   $  5_{\pm3}  -      4_{\pm3}     $    &        241832 &  84.6     & Y$^b$ & N\\
 CH$_3$OH-$A$   $  5_2       -      4_2          $    &        241842$^c$ &  72.5 & Y$^b$ & N\\
 CH$_3$OH-$E$   $  5_{3}     -      4_{3}        $    &        241843     &  82.5 & Y$^b$ & N\\
 CH$_3$OH-$E$   $  5_1       -      4_1          $    &        241879 &  55.9     & Y$^a$ & N\\
 CH$_3$OH-$E$   $  5_{\pm2}  -      4_{\pm2}     $    &        241904 &  $\sim 60$$^d$ & Y$^b$ & N\\
 \hline\hline

\end{tabular}
\begin{list}{}{}
\item $^a$ detected at the original resolution of the IRAM-30m telescope data
\item $^b$ detected at a resolution of $\sim 20''$
\item $^c$ blended with the $5_{3}-4_{3}$-$E$
\item $^d$ E$_{up}=$60.7~K for the $  5_{-2}- 4_{-2}$ line, 57.1~K for the $  5_{+2}- 4_{+2}$ transition.
\end{list}
\end{table}

The region mapped with the IRAM-30m telescope is presented in
Fig.~\ref{integrated}, where in the left panel we show the integrated
intensity of the strongest detected H$_2$CO line, while the right
panel shows the emission from the strongest CH$_3$OH transition along
the Orion Bar alone.  For comparison, the integrated intensity of the
H$^{13}$CN (1-0) from \citet{2003ApJ...597L.145L} is overlaid on our
data.  Both molecules clearly trace the molecular medium associated
with the Orion Bar, and peak to the southwest of it. This peak 
probably belongs to the low-intensity bridge that connects Orion South
to the Orion Bar. This structure was detected in previous observations
at velocities between 7 and 9~km~s$^{-1}$ \citep[e.g.,
][]{1995A&A...297..567T}. The southwest peak of CH$_3$OH and H$_2$CO
detected in our data has a peak velocity of 8~km~s$^{-1}$. In
addition, formaldehyde is detected towards the north of the map, at a
position that spatially coincides with continuum emission at
$350~\mu$m \citep{1998ApJ...509..299L} and with a velocity between 9
and 11 km~s$^{-1}$.

\begin{figure}
\centering
\subfigure[]{\includegraphics[angle=-90, width=7.5cm]{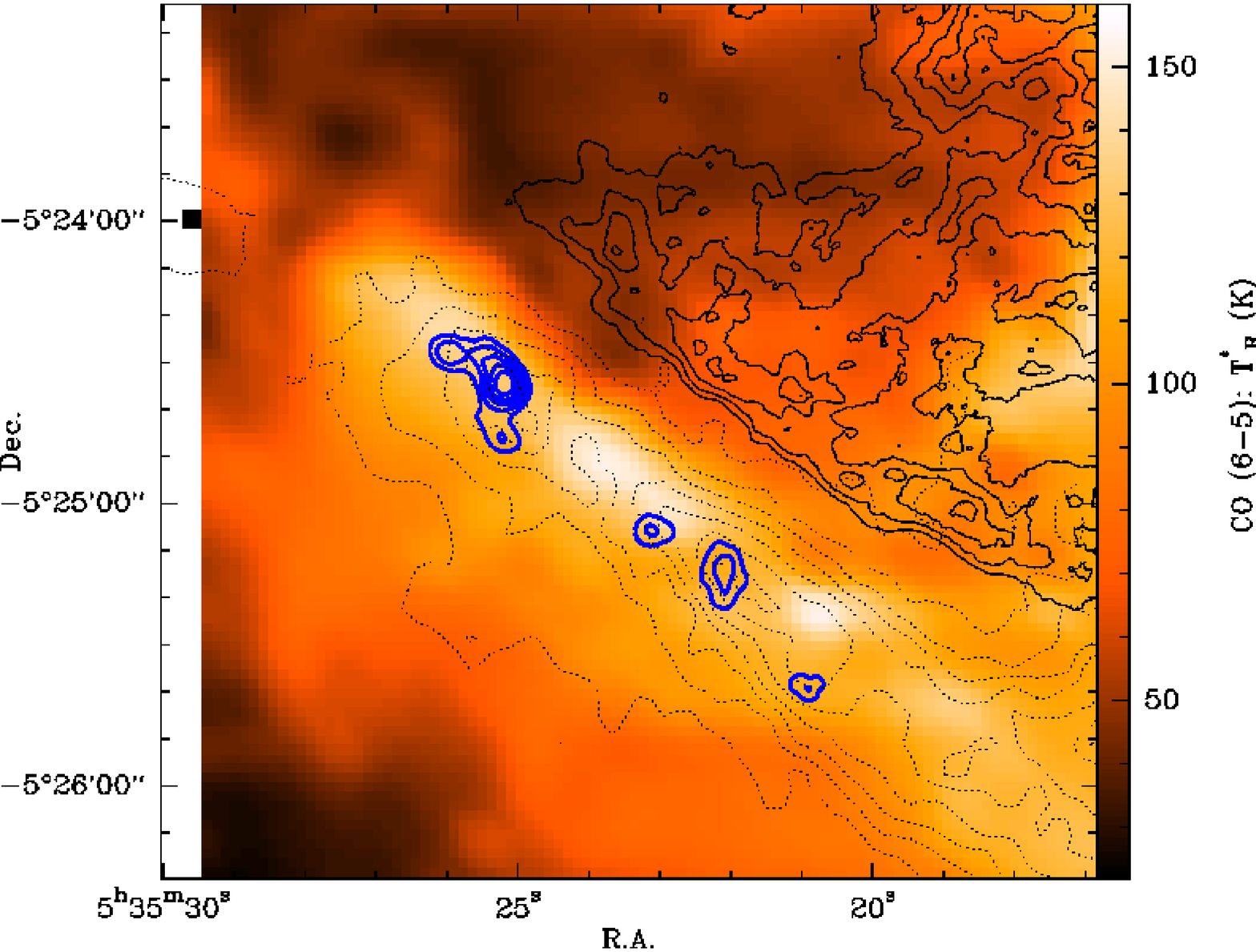}}
\caption{Distribution of the CO(6-5) peak brightness temperature
(colour image). The black contours (from 40$\%$ of the peak intensity,
in steps of 10\%) show the 20~cm continuum emission, which traces the
ionisation front.  The dotted contours (from 15$\%$ of the peak
intensity, in steps of 10\%) are the integrated intensity of the
H$_2$CO $3_{0,3}-2_{0,2}$~line; the blue contours (as in Fig.~\ref{integrated}) represent the dense clumps;  the square marks the position of the secondary peak identified in  H$_2$CO and CH$_3$OH.}\label{morphology}
\end{figure}

Along the Bar, both molecules trace the elongated region where
molecular clumps are located, with an absolute emission peak on clump
3 of \citet{2003ApJ...597L.145L}. Moreover, both species are detected
on a secondary peak, northeast of clump 1
($\alpha_{2000}=05^h35^m29.6^s, \delta_{2000}=-05^\circ23'59''.9$),
not covered by the observations of \citet{2003ApJ...597L.145L}, but
close to a peak of CO(6-5) (Fig.~\ref{morphology}), which could
represent an additional clump of dense gas along the Orion Bar.  To
better visualise the overall morphology of the Orion Bar, in
Fig.~\ref{morphology} we show the 20~cm continuum emission from
\citet{1990ApJ...361L..19Y}, which traces the
ionisation front; the CO(6-5) integrated emission \citep{1998ApJ...509..299L}, which shows the temperature distribution
of the molecular gas; and the H$^{13}$CN (1-0) emission, which traces the dense clumps.

\begin{figure}
\centering
\subfigure[]{
\includegraphics[angle=-90, width=8cm]{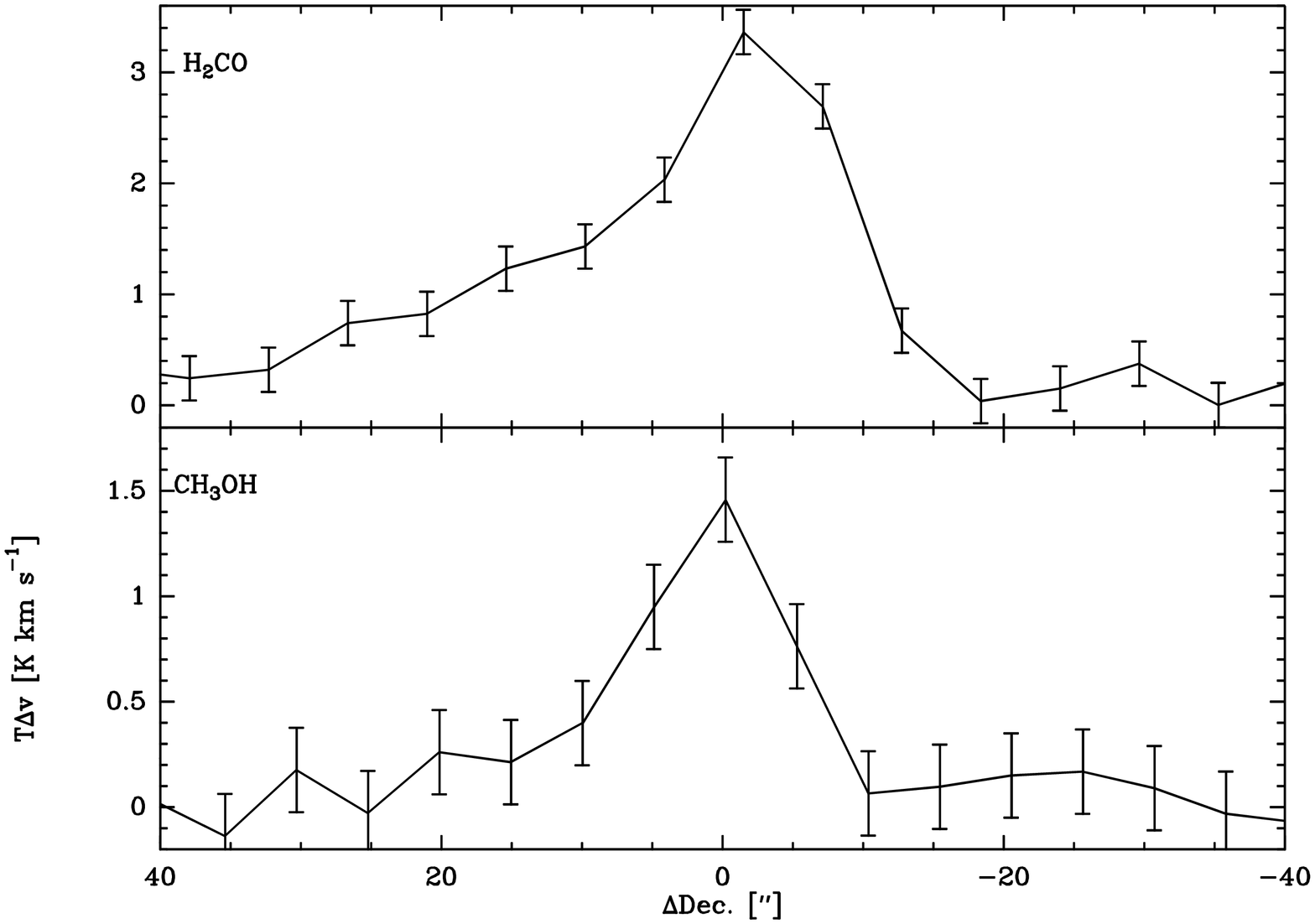}\label{strip-dec}}
\subfigure[]{
\includegraphics[angle=-90, width=8cm]{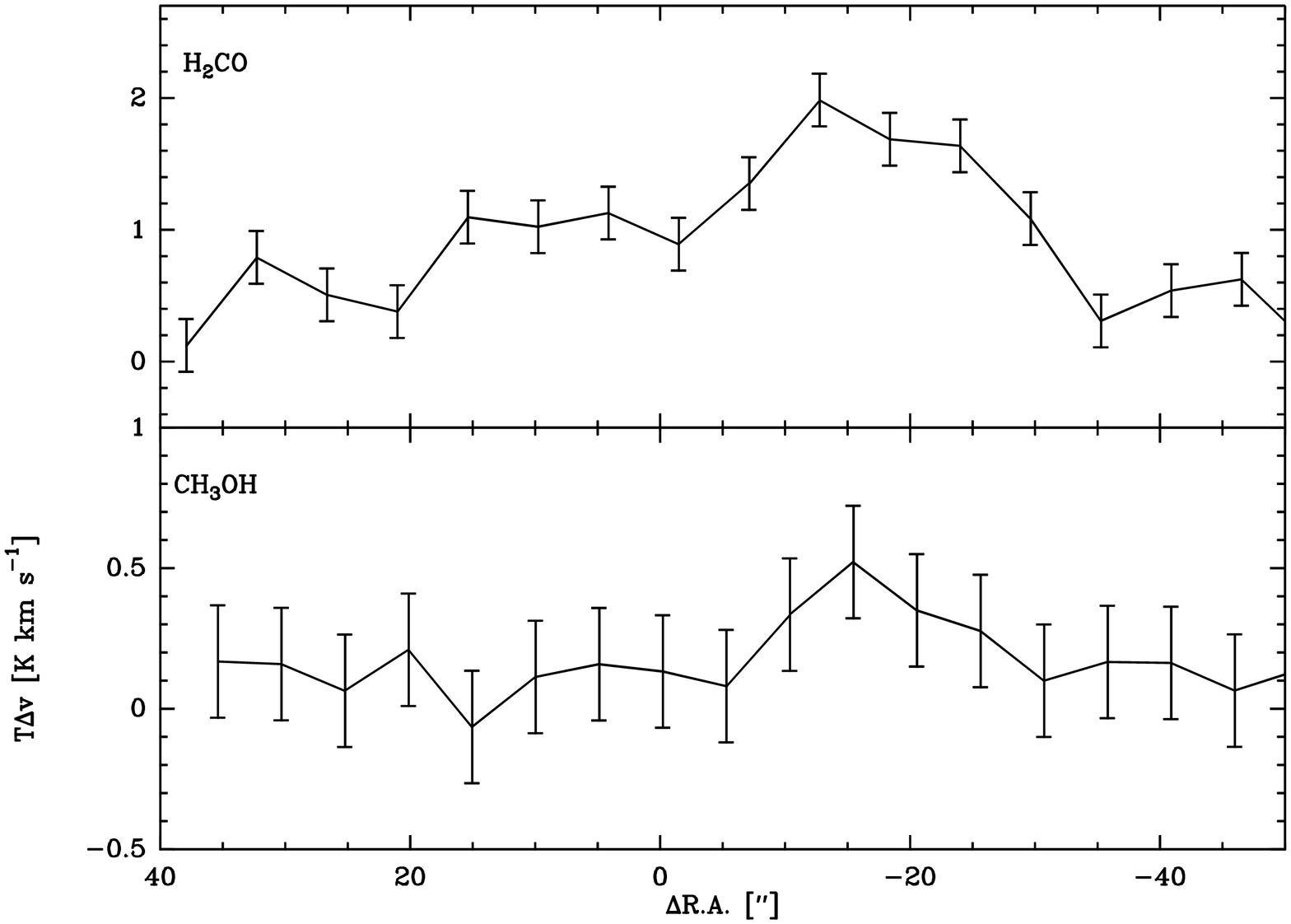}\label{strip-inter}}
\subfigure[]{
\includegraphics[angle=-90, width=8cm]{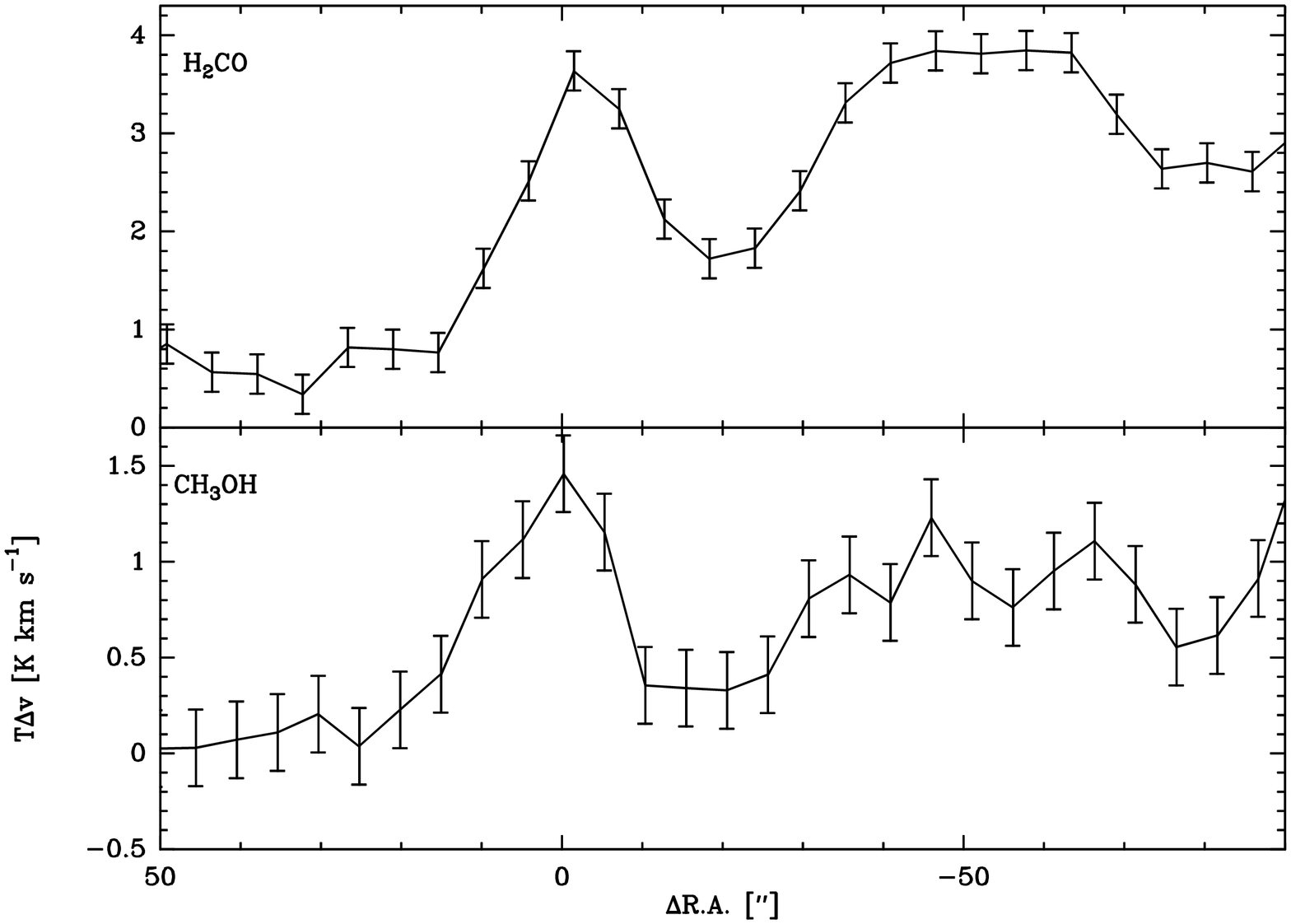}\label{strip-bar}}
\caption{Variation in the integrated intensity of the CH$_3$OH-$A$
($5_0-4_0$) and H$_2$CO $3_{2,2}-2_{2,1}$ lines observed with the IRAM-30m telescope across clumps 1 ({\bf a}), the interclump
medium between clump 1 and 2 ({\bf b}) and along the Bar ({\bf c}). In
Fig.~\ref{integrated} we outline the strips used to extract the
plots.}\label{strip-ra}
\end{figure}

In Fig.~\ref{spectra-h2co} and \ref{spectra-ch3oh} we compare the
H$_2$CO and CH$_3$OH spectrum towards clump 1 to the corresponding
spectra averaged over the region of the interclump gas outlined in
Fig.~\ref{integrated} (left).  The integrated intensity
of the CH$_3$OH $(5_0-4_0)$-$A$ transition in the interclump medium is a factor
$\sim 9$ lower than in the clump, the one of H$_2$CO
$3_{2,2}-2_{2,1}$~line a factor of 7 (see Table~\ref{lines}).  For a
better visualisation of the formaldehyde and methanol morphologies, we
show in Fig.~\ref{strip-ra} the variation in the integrated
intensities of the CH$_3$OH-$A$ ($5_0-4_0$) and H$_2$CO
$(3_{2,2}-2_{2,1})$ lines for three strips in the map; the exact
location of the strips is shown in Fig.~\ref{integrated}. The strip
across clump 1 (Fig.~\ref{strip-dec}) suggests a larger extension of
the H$_2$CO emission with respect to CH$_3$OH, while the variation in
the integrated intensity of the two species along the Bar is very
similar (Fig.~\ref{strip-bar}).  Although we present data for
the weakest of the two H$_2$CO lines, the signal-to-noise ratio in the
methanol data is still lower than for H$_2$CO and could bias our
results. From these strips and from the comparison between the
H$_2$CO emission and the CO(6-5) map, it also emerges that
formaldehyde extends to a greater distance from the clumps in the
molecular cloud than towards the ionisation front.  We also smoothed
the data to lower resolutions,  to verify whether the CH$_3$OH
and H$_2$CO emission are similarly affected by beam dilution. At a
resolution of 50$''$, the H$_2$CO emission is still $\sim 37\%$ of the
intensity of the original data, while at the same resolution the
CH$_3$OH peak intensity drops down to $\sim 18\%$ of the original
value.  These tests suggest that formaldehyde emission is associated
with the clumps but also with the interclump medium, while methanol
emission is only found  in the dense molecular clumps.  In Sect.~\ref{discussion}, we investigate whether the different
morphologies of the two molecular species stem from an excitational
or observational bias, or whether they imply different abundances of
H$_2$CO and CH$_3$OH in the two media.

\begin{figure}
\centering
\includegraphics[angle=-90, width=8.5cm]{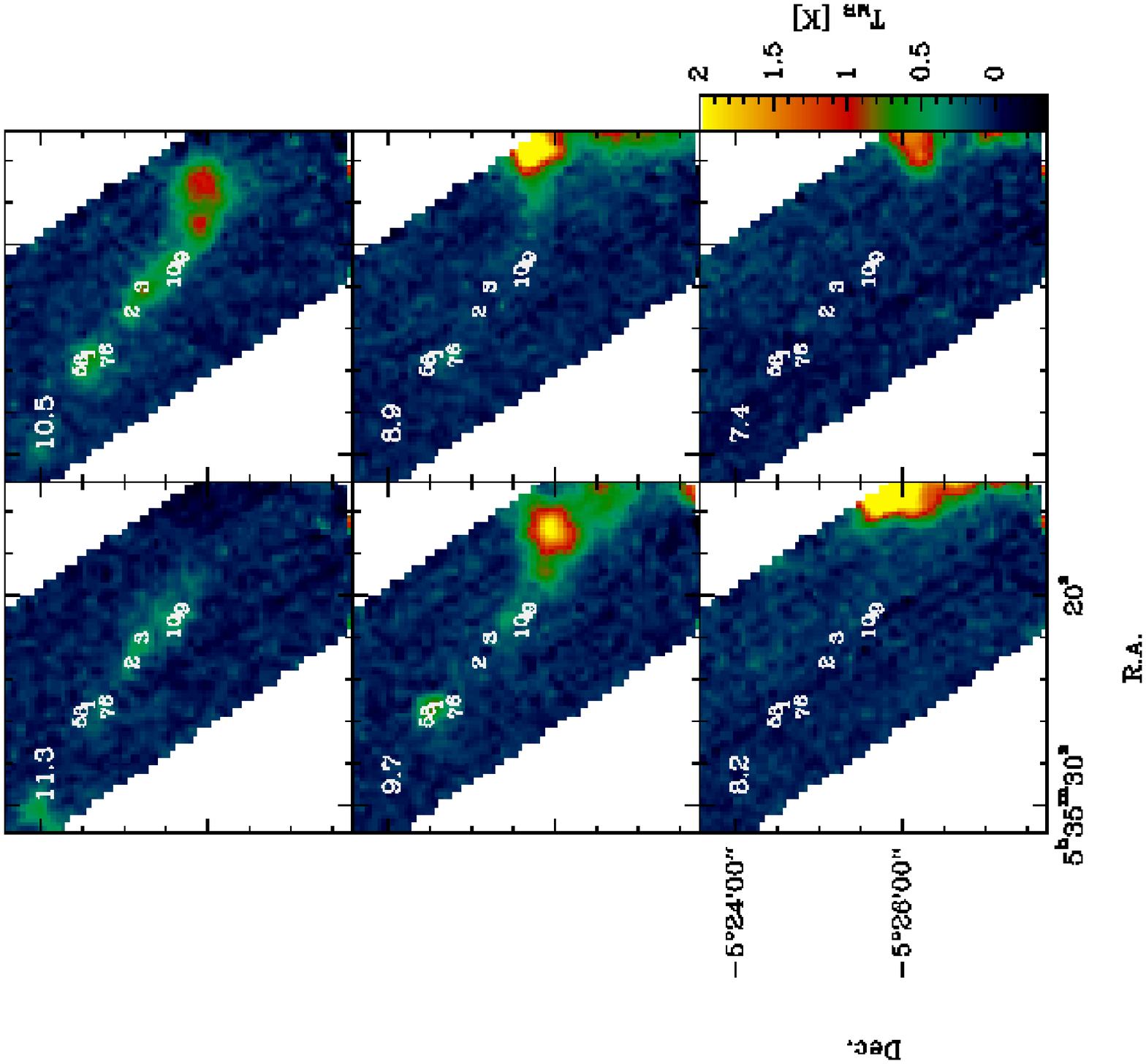}
\caption{Channel maps of the CH$_3$OH-$A$ ($5_0-4_0$) emission between
11.3 and 7.4 km~s$^{-1}$   observed with the IRAM-30m telescope. 
}\label{ch3oh-chanmaps}
\end{figure}

Finally, the spectral resolution used for the methanol observations
allows us to study the velocity field along the Bar. The channel maps
of the $(5_0-4_0)$-$A$ line (Fig.~\ref{ch3oh-chanmaps}, smoothed to
0.77~km~s$^{-1}$) show that clump 1 peaks around 9.5~km~s$^{-1}$ and
clump 3 around 10.5, as for H$^{13}$CN
\citep{2003ApJ...597L.145L}. The secondary peak detected to the
northeast of clump 1 has a peak velocity around 11~km~s$^{-1}$, while
the molecular cloud to the south of the Bar is red-shifted with
respect to the clumps ($v_{\rm{lsr}}\sim 8$~km~s$^{-1}$). This trend
in velocity (blue-shifted velocities in the northeast, red-shifted
values in the southwest) was also found by
\citet{2000ApJ...540..886Y} in their HCN maps.

\begin{table*}
\centering
\caption{Line parameters\label{lines}}
\begin{tabular}{lrrrlrr}
\hline
\hline
&\multicolumn{1}{c}{$T_{\rm{MB}}$}&\multicolumn{1}{c}{$\int T_{\rm{MB}}dv$}&\multicolumn{1}{c}{$\Delta v^a$}&\multicolumn{1}{c}{$T_{\rm{MB}}$}&\multicolumn{1}{c}{$\int T_{\rm{MB}}dv$}&\multicolumn{1}{c}{$\Delta v^a$}\\
&\multicolumn{1}{c}{[K]}&\multicolumn{1}{c}{[km~s$^{-1}$]}&\multicolumn{1}{c}{[km~s$^{-1}$]}&\multicolumn{1}{c}{[K]}&\multicolumn{1}{c}{[K~km~s$^{-1}$]}&\multicolumn{1}{c}{[km~s$^{-1}$]}\\
&\multicolumn{3}{c}{H$_2$CO ($3_{0,3}-2_{0,2}$)}&\multicolumn{3}{c}{CH$_3$OH ($5_0-4_0$-$A$)}\\
\hline
clump~1&3.2&8.9&2.0&1.1&1.8&1.4\\
interclump&0.3&1.2&3.1&0.06&0.2&4.0\\
\hline
\end{tabular}
\begin{list}{}{}
\item $^a$ corrected for the spectral resolution, $\Delta v=\sqrt{(\Delta v_{\rm obs})^2-(\Delta v_{\rm res})^2}$
\end{list}
\end{table*}

\subsection{Hybrid interferometric+single-dish data}\label{hybrid}

Figure~\ref{fig_cubes} compares the 30m (top panels) and hybrid 30m+PdBI
(bottom panels) data cubes. The left column displays the images integrated
between 8.3 and 11.7 km/s. The middle column displays the spectra averaged
over the whole interferometric field-of-view.  The hybrid spectra are identical to the
  30m ones within the noise level, confirming that all the 
extended emission filtered out by the PdBI is correctly recovered from the single-dish data. 
The right column of Fig.~\ref{fig_cubes} displays the spectra averaged over the central clump
marked in the left column. The hybrid spectrum is brighter
than the 30m spectrum implying beam dilution in the single-dish data.

From the PdBI data and the hybrid 30m+PdBI data, we can therefore conclude that the H$_2$CO emission
 comes from both an extended component and  a compact one, still unresolved at the resolution of the IRAM-30m telescope.

\section{Discussion}\label{discussion}
\subsection{Formaldehyde}\label{disc-h2co}

Line ratios of formaldehyde lines can be used to infer the physics of
the gas in dense molecular regions \citep{1993ApJS...89..123M}.
Because of the coarse spectral resolution of our observations ($\Delta
v\sim 1.7$~km~s$^{-1}$), the line profiles of the H$_2$CO transitions
are poorly resolved. As a result, we do not have a correct measure of the
peak intensity of the lines, diluted over the beam, but our values are a lower limit to the
true main-beam line intensities.  However, the areas are conserved quantities
and can be used to study the excitation of H$_2$CO in the Orion Bar.

Figure ~\ref{ratio} shows the ratio of the integrated intensities of
the $3_{0,3}-2_{0,2}$ line to the $3_{2,2}-2_{2,1}$ line. The analysis
is limited to the region where the signal-to-noise ratio of the
$3_{2,2}-2_{2,1}$ integrated intensity is at least equal to 3.  The
ratio increases from the northeast, where the clumps are, to the
southwest. 

  \begin{figure*}
    \centering{}
     \includegraphics[height=\hsize,angle=270]{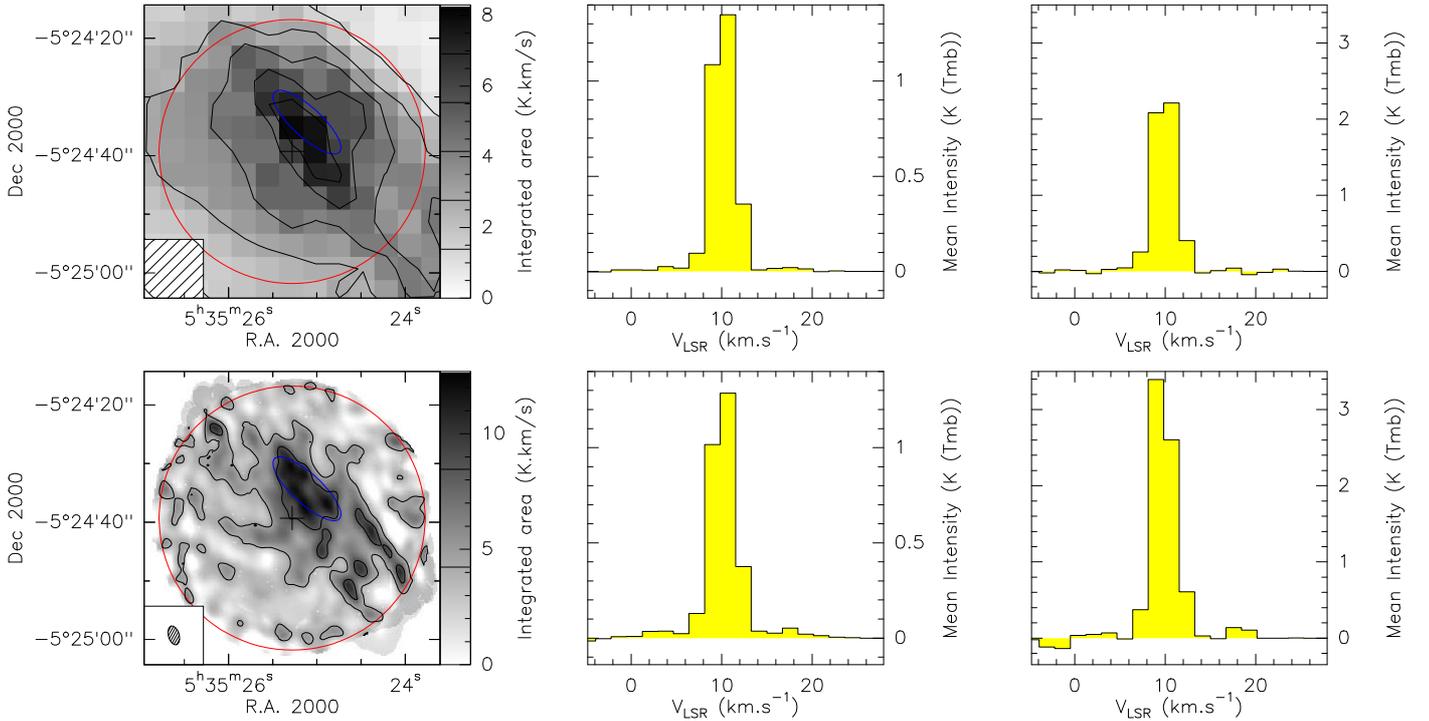}
    \caption{\emph{Top line:} IRAM-30m data. \emph{Bottom line:} hybrid
      30m+PdBI deconvolved data cube. \emph{Left column:} Images of the
       integrated line intensity of the H$_2$CO ($3_{0,3}-2_{0,2}$) transition between 8.3 and 11.7 km/s. The values of the
      contour levels are shown on the colour scale. \emph{Middle column:}
      Spectra averaged over the whole interferometric field-of-view marked
      as the red circle on the images. \emph{Right column:} Spectra
      averaged over the central clump marked with the blue ellipse on the
      images.}
    \label{fig_cubes}
  \end{figure*}

We ran LVG calculations for column densities of H$_2$CO between
10$^{12}$ and 10$^{15}$~cm$^{-2}$, temperatures in the range 20--200~K,
and densities of 10$^4$--10$^8$~cm$^{-3}$, and calculated the behaviour
of the $\chi^2$ value of the observed $3_{0,3}-2_{0,2}$ and
$3_{2,2}-2_{2,1}$ line intensities as function of these quantities.  Two possible
regimes are found (see Fig. \ref{chisq}), independent of
temperature. At low densities (n$<10^6$~cm$^{-3}$), the $\chi^2$ value
depends only on the product of the H$_2$ density and H$_2$CO column
density (low-density branch). For higher densities, the $\chi^2$ value
varies only slowly with H$_2$ density and is mostly a function of
H$_2$CO column density (high-density branch).

 For the values in the interclump gas, assuming a beam filling factor
 of unity, we find a nominal minimum for
 the interclump medium at $T=180$~K, n$(\rm{H}_2)$ = $6.6 \times
 10^5$~cm$^{-3}$, $N_{\rm {H_2CO}} = 1.4 \times
 10^{13}$~cm$^{-2}$. However, as  can be seen in Fig. \ref{chisq},
 these parameters are not well constrained. Using the $1\sigma$
 confidence level, we derived a lower limit to the kinetic temperature
 of the interclump equal to 76~K, and the product $n_{\rm{H}_2} \times
 N_{\rm H_2CO}$ is equal to $\sim 3\times 10^{18}$~cm$^{-5}$ for
 densities smaller than 10$^6$~cm$^{-3}$. By comparison with PDR
 models, \citet{2000ApJ...540..886Y} found densities of hydrogen
 nuclei of 10$^4$-$10^5$~cm$^{-3}$ in the interclump; similarly,
 \citet{1995A&A...294..792H} found an interclump H$_2$ density of
 $3\times 10^4$~cm$^{-3}$ by modelling mm line observations.  Thus,
 by using a molecular hydrogen density between $5\times 10^3$ and
 $5\times 10^4$~cm$^{-3}$ we can constrain the column density of
 H$_2$CO to values in the range $6\times 10^{13}-6\times
 10^{14}$~cm$^{-2}$.  The high-density branch does not appear to be
 relevant for the interclump medium.

In Sect.~ \ref{hybrid} we analysed the combined PdBI+30m data and 
concluded that the H$_2$CO emission is composed of two components, one extended and one unresolved at the resolution 
of the 30m telescope at this frequency. However from these data,
  it is difficult to infer the exact size of clump 1 and the contribution of
the interclump gas to the total emission of H$_2$CO. In Fig.~\ref{fig_cubes}, we marked clump 1 with an ellipse of 
$15''\times 5''$ (P.A.=47$^\circ$) and showed that the IRAM-30m data are affected by beam dilution at this scale, implying a smaller size for the clump.
Therefore, for clump 1, we corrected the main-beam line intensities for the
contribution from the interclump and for the beam dilution assuming a
typical size for the clump of $7''$ \citep[see
][]{2003ApJ...597L.145L}.  As expected given the similar value of the
line ratio on clump 1 and in the interclump, the values derived are
similar to those of the interclump: $T=170$~K, n$(\rm{H}_2)$  = $6.6
\times 10^5$~cm$^{-3}$, $N_{\rm {H_2CO}} = 1.6 \times
10^{14}$~cm$^{-2}$. The $1\sigma$ contour gives a value of $3.8\times
10^{19}$~cm$^{-5}$ for the product $n_{\rm{H}_2} \times N_{\rm H_2CO}$
for $N_{\rm {H_2CO}} >5 \times 10^{13}$~cm$^{-2}$ and $n(\rm{H}_2) <
5\times 10^5$~cm$^{-3}$.  For the low-density branch, we derive a
lower limit of 50~K for the kinetic temperature. For the high-density
branch, more appropriate for the clump, the column density is
constrained to values $2.8\times 10^{14}-5.4 \times
10^{14}$~cm$^{-2}$. A lower limit of 80~K can be derived for the
kinetic temperature.  It may appear surprising that the temperature is
so little constrained, although the ratio of these lines has been used
as a temperature tracer
\citep{1993ApJS...89..123M,1995A&A...294..792H}.  However, for the
values of the line ratio we observe, the LVG predictions show a more
complex behaviour with temperature \citep[see also Fig.~13
of][]{1993ApJS...89..123M}.  Taking the measurement errors into
account by calculating the $\chi^2$ value, as we do, and using the
line intensities and not just the line ratios then constrains only a
lower limit to the temperature.

The results of the LVG analysis of H$_2$CO are presented in Table~\ref{results}.
\begin{figure}
\centering \includegraphics[angle=-90, width=8cm]{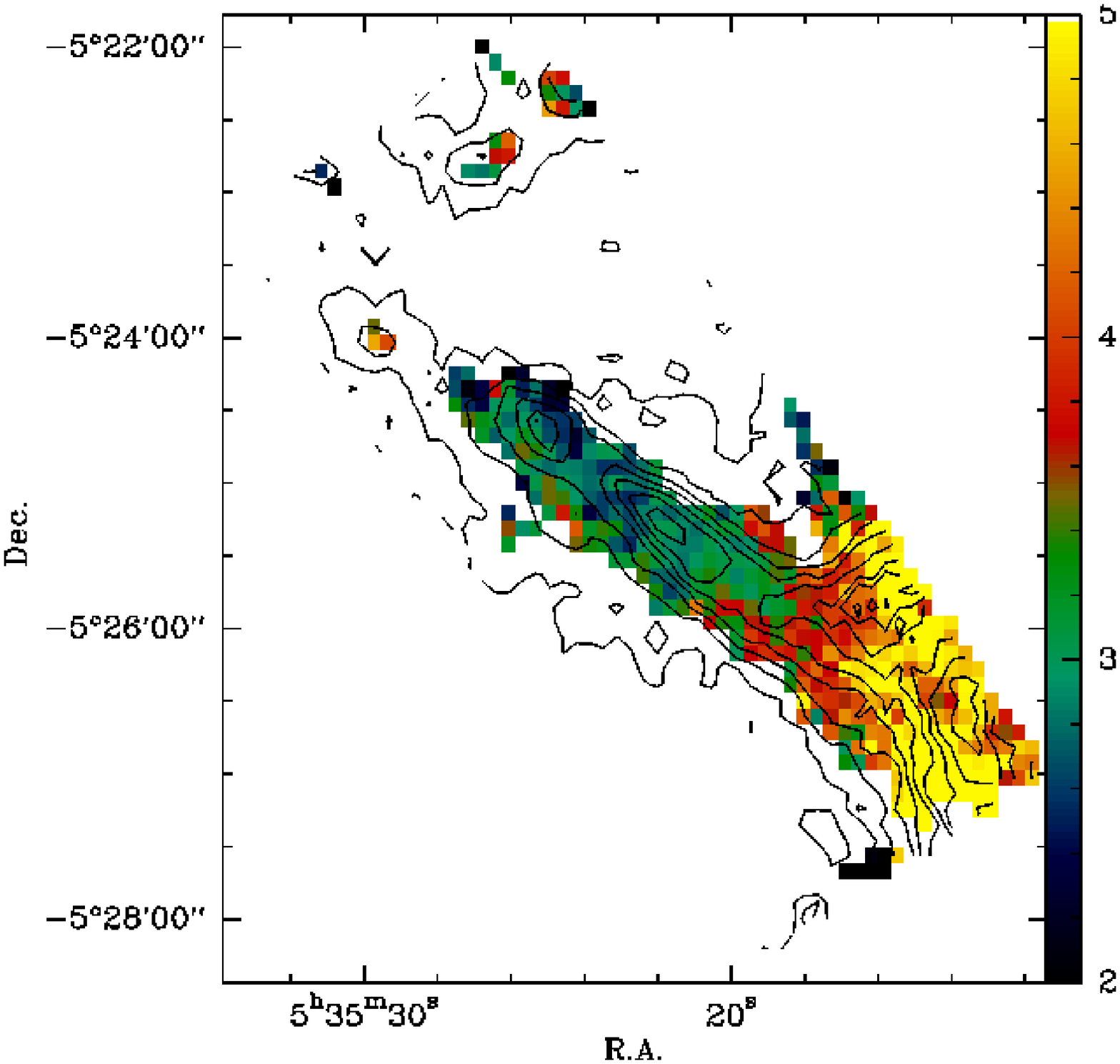}
\caption{Ratio of the integrated intensity of the H$_2$CO
$3_{0,3}-2_{0,2}$ transition to the $3_{2,2}-2_{2,1}$. The solid black
contours show the integrated intensity of the $3_{0,3}-2_{0,2}$ line
(levels are from 1.8~K~km~s$^{-1}$ in steps of 1.8).  Typical errors along the Bar are of the order of 0.3.}\label{ratio}
\end{figure}

\begin{figure}
\centering
\includegraphics[bb= 34 207 536 588,clip,width=8cm]{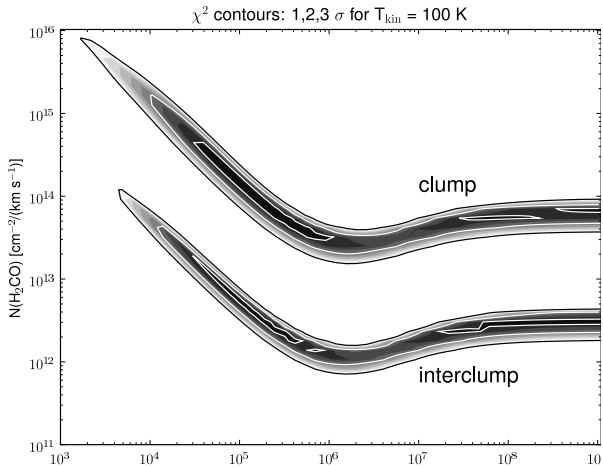}
\caption{$\chi^2$ square distribution of the  $3_{0,3}-2_{0,2}$ and $3_{2,2}-2_{2,1}$ line intensities in the [$n$(H$_2$), $N({\rm
H_2CO})$] plane for a temperature of 100~K for the two sets of
observations clump and interclump shown in Fig.~\ref{spectra-h2co}. The plot shows, for
each spatial region, the low-density branch and the high-density
branch discussed in the text. Note that to derive the column
densities given in \S~\ref{disc-h2co}, one has to multiply with the
line widths.}\label{chisq}
\end{figure}
\begin{table*}
\centering
\caption{Summary of the model results\label{results}}
\begin{tabular}{rccc}
\hline
\hline
&\multicolumn{1}{c}{$N_{{\rm H_2CO-}p}$}&\multicolumn{1}{c}{$N_{{\rm CH_3OH-}A}$}&\multicolumn{1}{c}{${[\rm CH_3OH}-A]/[{\rm H_2CO}-p]$}\\
&\multicolumn{1}{c}{[$10^{13}$~cm$^{-2}$]}&\multicolumn{1}{c}{[$10^{13}$~cm$^{-2}$]}&\multicolumn{1}{c}{}\\
\hline
clump~1&$(28-54)^a$&$23-31$&$0.4-1.1$\\
interclump &$(6-60)^b$&$(0.04-4)^b$&$7\times 10^{-4}-0.7$\\
&&$(0.04-0.4)^c$&$(7\times 10^{-4}-7\times 10^{-2})^c$\\
\hline
\end{tabular}
\begin{list}{}{}
\item $^a$ assuming $n=10^6$~cm$^{-3}$
\item $^b$ assuming $n=(5\times 10^3-5\times 10^4)$~cm$^{-3}$
\item $^c$ assuming $T>50$~K 
\end{list}
\end{table*}

\subsection{Methanol}\label{disc-ch3oh}
\citet{2004A&A...422..573L} have studied the excitation of the CH$_3$OH
($5_k-4_k$) band as function of the temperature, density, and methanol
column density of the gas for a range of physical conditions applicable to  the Orion Bar. 
They concluded that line ratios within this band are
not sensitive to the temperature of the gas for temperatures higher
than 30~K, but only depend on the density.

The methanol emission from clumps 1 and 3 is studied in a separate
paper \citep{paper_i}: clump 1 is found to be warmer than
clump 3, although the errors are large ($3\sigma$ ranges: $T_1\sim
45^{+47}_{-17}$~K, $T_3\sim 35^{+17}_{-15}$~K), but with similar
column densities and densities ($N_{\rm CH_3OH}\sim 3 \times
10^{14}$~cm$^{-2}$, $n\ge 5\times 10^6$~cm$^{-3}$). These values were
derived by modelling the $5_k-4_k$ and $6_k-5_k$ bands observed with
the IRAM-30m (observations presented in this paper) and the APEX
telescopes, respectively. For the models, the IRAM data were smoothed
to the resolution of the APEX data ({\it FWHM} beam $\sim 20''$). A
source size of $10''$ was used, since several clumps fall in the beam
of the observations. This is equivalent to assuming that all clumps have
similar physical conditions. Since the CH$_3$OH emission was found to
be optically thin, we can correct the results for a source size of $7''$ (as assumed for the model of H$_2$CO) and 
derive the equivalent column density. This
corresponds to a total column density of $5.4 \times
10^{14}$~cm$^{-2}$, or to $2.7 \times 10^{14}$~cm$^{-2}$ for
CH$_3$OH-$A$, assuming that the two symmetric states have the same
column density.  The $1\sigma$ confidence level is $2.3\times
10^{14}-3.1\times 10^{14}$~cm$^{-2}$.

\begin{figure}
\centering
\includegraphics[angle=-90, width=8cm]{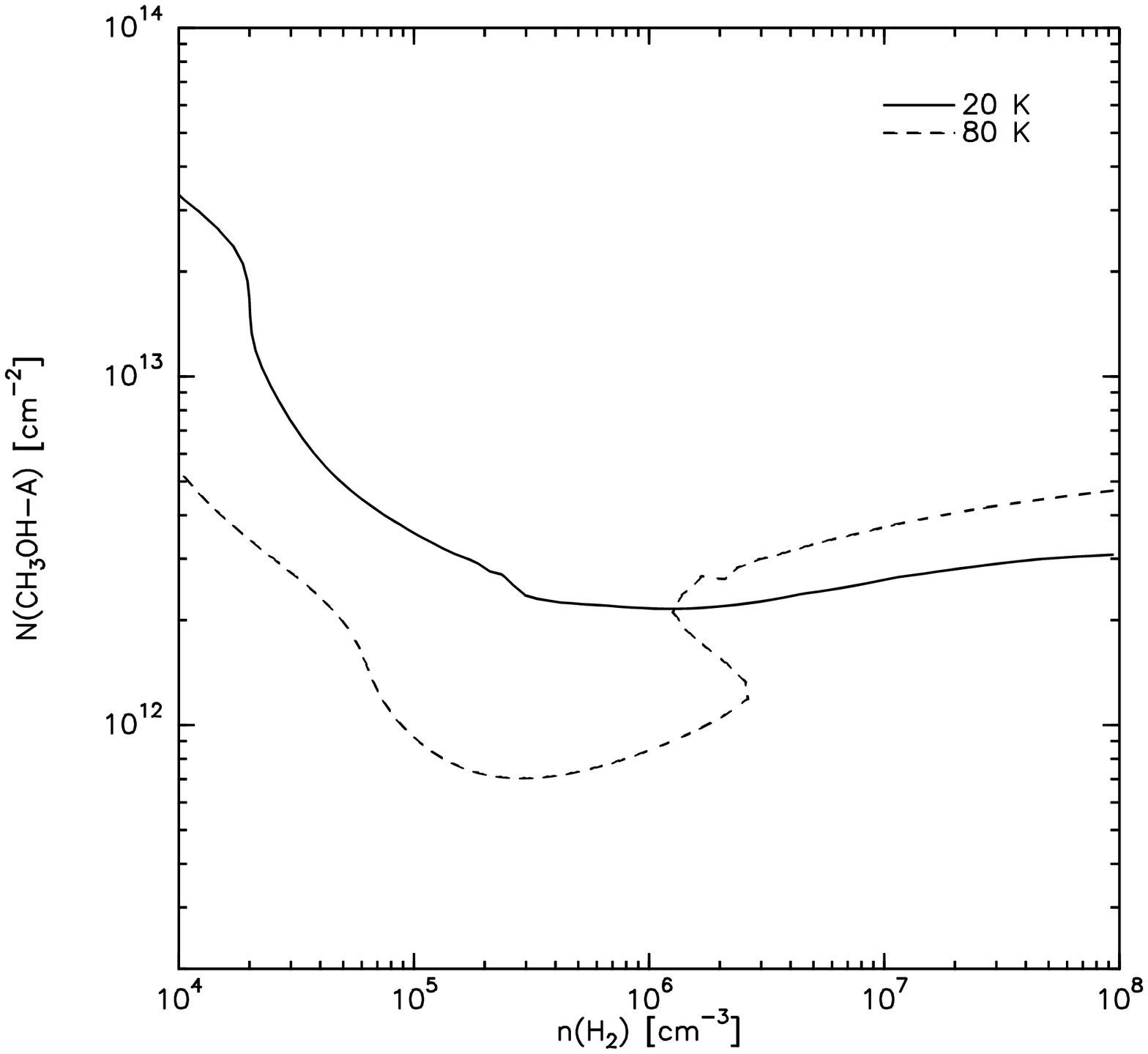}
\caption{Results of the statistical equilibrium calculations for CH$_3$OH. The
solid and dashed lines show the
line intensity of the $5_0-4_0$-$A$ line as measured towards the interclump medium (0.06~K) for temperatures of  20 and 80~K, 
respectively,
as functions of density and column density of CH$_3$OH-$A$ .}\label{colden}
\end{figure}

For the interclump medium, we analysed the CH$_3$OH spectrum averaged
over the area of the interclump medium outlined in
Fig.~\ref{morphology}.  We ran models for the excitation of methanol
in the range of densities $10^4-10^8$~cm$^{-3}$, column densities
$10^{12}-10^{18}$~cm$^{-2}$, and temperatures $20-200$~K.  The observed
main-beam brightness temperature of the ($5_0-4_0$)-$A$ line in the
interclump medium is $0.06\pm 0.01$~K. Assuming that the emission
comes from an extended source (e.g., beam filling factor $\eta_c=1$),
we find a situation similar to the one described for H$_2$CO and we can
identify two different regimes. In the first ($n<10^6$~cm$^{-2}$), a
fit to the data is found for $N_{{\rm CH_3OH}-A}<1.3 \times
10^{14}$~cm$^{-2}$: the column density decreases with increasing
density, but their product is almost constant (between $ 2\times
10^{17}$~cm$^{-5}$, $50~\rm{K}\le T<200~\rm{K}$, and $ 2\times
10^{18}$~cm$^{-5}$, $20~\rm{K}<T<50~\rm{K}$).  For higher densities,
the $5_0-4_0$ line thermalises. In this regime, the column density
increases slowly ($N_{{\rm CH_3OH}-A}=1.3\times 10^{12}-2 \times
10^{13}$~cm$^{-2}$).  For both regimes, the dependence on the kinetic
temperature is not strong and the results presented are valid for
temperatures in the range 20--200~K. However, increasing the
temperature of the gas implies a decrease in the column density of
methanol for the low-density case ($T\ge 50$~K, $N_{{\rm
CH_3OH}-A}<3\times 10^{13}$~cm$^{-2}$, see Fig.~\ref{colden}).

As already discussed in the previous section, from the literature we
know that the density of the interclump gas is less than
$10^6$~cm$^{-3}$, and therefore that the low-density regime applies to
the modelling of the interclump. The best fit to the data is found for
$N_{{\rm CH_3OH-}A}\sim 3 \times 10^{12}$~cm$^{-2}$, $T\sim 80$~K,
$n\sim 9\times 10^4$~cm$^{-3}$.  For densities in the range $5\times
10^3-5\times 10^4$~cm$^{-3}$, and for the values of the product
$N\times n$ given above, we derive column densities in the range
$4\times 10^{11}-4\times 10^{13}$~cm$^{-2}$. However, if we allow only
temperatures above 50~K, as found in our data and by other authors,
the inferred methanol column density in the interclump is $4 \times
10^{11} \le N \le 4 \times 10^{12}$~cm$^{-2}$ (see Table~\ref{results}).

\section{Abundances of methanol and formaldehyde in the Orion Bar}

From the analysis performed in the previous section, we can compute
the abundance of methanol relative to formaldehyde in the interclump
medium and in the dense clumps, and verify that the drop of
intensity of the methanol emission in the interclump medium is the result of
the different physics of the interclump relative to the dense clumps,
or whether it reflects a real decrease in the CH$_3$OH abundance.

Given an estimate of H$_2$ column density, we could also infer the
abundance of both molecules relative to molecular hydrogen. An
estimate of the H$_2$ column density can be derived from the continuum
emission at (sub)millimetre wavelengths, assuming a given temperature
for the dust and optically thin emission.  The continuum emission at
350~$\mu$m of the Orion Bar was studied by
\citet{1998ApJ...509..299L}. These authors defined an average
temperature for the dust in the OMC-1 region of 55~K on the basis of
its FIR colours \citep{1976ApJ...204..420W} and excluded colder
temperatures for the Orion Bar given the good agreement of the
350~$\mu$m continuum emission and the CO(6-5) emission, which
originated in the outer PDR layers. 
However,
\citet{1976ApJ...204..420W} found an average FIR colour temperature of
75~K in the Orion Bar, which could imply that the Orion Bar is
significantly warmer than the rest of OMC-1. 
For a dust temperature
of 55~K in the interclump medium, and for the mean value of the
350~$\mu$m emission over the region used to derive the average spectra
of the interclump medium (Fig.~\ref{integrated}), we derive a column
density of H$_2$ of $5.3\times 10^{21}$~cm$^{-2}$. This translates into
to $3.4\times 10^{21}$~cm$^{-2}$ for $T_d =75$~K. A value of
0.101~g~cm$^{-2}$ was used for the dust opacity
\citep{1994A&A...291..943O}. For clump 1,
for a temperature of 45~K (as derived from methanol), and under the assumption that   the dust and the gas are
thermally coupled at
the high density of the clumps \citep[e.g.,][]{1984A&A...130....5K}, the H$_2$
column density is $2\times 10^{23}$~cm$^{-2}$. This is not the average
column density on the beam, but it was corrected for a source size of
$7''$ and for the contribution of the interclump medium at the
position of clump 1. These values translate into abundances of H$_2$CO
in the range $(11-110)\times 10^{-9}$ for the interclump,$(1.4-2.7)\times 10^{-9}$ for the clumps.  For methanol, we obtain an
abundance relative to H$_2$ of $(0.08-8)\times 10^{-9}$ for the
interclump, $(1.2-1.6)\times 10^{-9}$ for the clumps. 

However, these estimates are affected by large uncertainties: 
a difference of 20~K in
the dust temperature implies an uncertainty of a factor of $\sim 2$ in the
estimate of the H$_2$ column density in the interclump, uncertainty
that could be even greater if the average temperature in the region of
the interclump used in our analysis is higher than 75~K.  Similarly,
the estimate of the H$_2$ column density on the clumps is affected by
large uncertainties because of the assumption that the dust is thermally
coupled to the gas at high densities and because of the large errors on the
temperature of the gas from the methanol spectrum ($28<T_1<92$~K,
$20<T_3<52$~K). 

On the other hand, the uncertainties in the determination of the
abundance of CH$_3$OH relative to H$_2$CO are related only to the
errors in the estimate of the column density of the two molecules,
which already include the uncertainties in the temperature.
Therefore, we believe $[{\rm CH_3OH}-A]/[{\rm H_2CO}-p]$ to be more
solidly estimated than $[{\rm CH_3OH}-A]/[{\rm H_2}]$ or $[{\rm
H_2CO}-p]/[{\rm H_2}]$.
Using the column densities derived in the previous section, we infer
an abundance of CH$_3$OH relative to H$_2$CO of the order 0.4--1.1 in
the dense clumps, and 7$\times 10^{-4}$-- 7$\times 10^{-1}$  in the interclump. If we
assume that the temperature of the interclump medium is higher than
50~K, then abundance of methanol drops down to 7$\times
10^{-4}$-- 7$\times 10^{-2}$.  Since it is a reasonable assumption that
the temperature of the interclump medium is 50~K or even higher, we conclude
that the abundance of methanol relative to formaldehyde decreases by
at least one order of magnitude in the interclump medium in comparison
to the dense clumps. 

A possible explanation for the decrease in abundance of methanol relative to formaldehyde from the clumps
 to the interclump medium is photodissociation in a low-density enviroment. Although the photodissociation 
rates of CH$_3$OH and H$_2$CO are of the same order of magnitude \citep{2000A&AS..146..157L}, one of the products
of the photodissociation of methanol is formaldehyde. Therefore, both molecules can be destroyed 
by the radiation field in the interclump medium, but H$_2$CO may be
 replenished through the photodissociation of CH$_3$OH. 
However, large uncertainties affect the measurement
of the photodissociation rates, and so more accurate models would be required to test this hypothesis. Alternatively, as already
discussed by \citet{2006A&A...454L..47L}, H$_2$CO could be formed in the interclump medium through gas phase reactions, which are not
efficient for the formation of CH$_3$OH \citep{2002luca}.

\section{Conclusion}
In a previous analysis of methanol and formaldehyde towards the Orion
Bar, we suggested that the two molecules might trace different
environments and that, while CH$_3$OH is associated with the clumpy
molecular cores of the Bar, H$_2$CO is found in the interclump
material.  To test this hypothesis, we mapped the Orion Bar in both
molecular species with the IRAM-30m telescope, with a spatial
resolution slightly higher than the expected size of the
clumps. Additional data were taken with the IRAM Plateau de Bure
Interferometer in H$_2$CO.

Both molecules are detected in the Orion Bar in our single-dish
data. Our data show that CH$_3$OH peaks towards the clumps of the Bar,
but its intensity decreases below the detection threshold in the
interclump at individual positions. By averaging over a large region
of the interclump medium, the strongest of the CH$_3$OH lines in our
setup ($5_0-4_0$-$A$) is detected with a peak intensity of $\sim
0.06$~K.
On the other hand, contrary to the hypothesis formulated in our previous study, formaldehyde is detected towards the clumps and the
interclump gas. 

Using an LVG program, we studied the excitation of H$_2$CO and CH$_3$OH
in the Orion Bar.  We suggest that formaldehyde is present in both
components of the Bar (clumps and interclump material), with column
densities up to $5.4\times 10^{14}$ ~cm$^{-2}$ in the clumps and between $6\times
10^{13}-6\times 10^{14}$~cm$^{-2}$ in the interclump. These values only
refer to para-formaldehyde.  From the analysis of methanol, we
concluded that the reason for the drop of intensity of CH$_3$OH in the
interclump medium is not the different physical conditions
with respect to the clumps, but a real drop in its column density
compared to the clumps (down to $4\times 10^{11}$~cm$^{-2}$ for
temperatures above 50~K). The abundance of methanol relative to
formaldehyde decreases by at least one order of magnitude in the
interclump medium compared to that in the dense clumps.

Despite the large errors on our estimates, our observations reveal
that the column density of methanol and formaldehyde in the clumps are
of the same order of magnitude, while the column density of methanol
in the interclump is lower than that of formaldehyde.   
This may be a result of photodissociation
of CH$_3$OH in the unshielded interclump gas, or it may reflect 
 that H$_2$CO can be produced in the gas phase more efficiently
than CH$_3$OH.  
Definitive observational conclusions would require a
much deeper integration towards the interclump gas and a better
characterisation of its density and temperature. More
detailed chemical models of PDRs, including 
grain surface reactions, are clearly needed to properly interpret
these observational results.

\acknowledgements{We are grateful to Helmut Wiesemeyer for his support before and during the observations,
and to the observers who obtained part of the data presented in this work during  pooled observations at the IRAM-30m telescope.}

\bibliographystyle{aa}
\bibliography{biblio_leu}

\end{document}